\documentclass[preprint,10pt]{aastex}

\begin{document}
  
  \title{Virial Masses of Black Holes from Single Epoch Spectra of
  AGN}

  \author{Brandon C. Kelly and Jill Bechtold}
  \email{bkelly@as.arizona.edu, jbechtold@as.arizona.edu}
  \affil{Steward Observatory, University of Arizona, 933 N Cherry Ave,
  Tucson, AZ 85721}
  
  \begin{abstract}

    We describe the general problem of estimating black hole masses of
    AGN by calculating the conditional probability distribution of
    $M_{BH}$ given some set of observables. Special attention is given
    to the case where one uses the AGN continuum luminosity and
    emission line widths to estimate $M_{BH}$, and we outline how to
    set up the conditional probability distribution of $M_{BH}$ given
    the observed luminosity, line width, and redshift. We show how to
    combine the broad line estimates of $M_{BH}$ with information from
    an intrinsic correlation between $M_{BH}$ and $L$, and from the
    intrinsic distribution of $M_{BH}$, in a manner that improves the
    estimates of $M_{BH}$. Simulation was used to assess how the
    distribution of $M_{BH}$ inferred from the broad line mass
    estimates differs from the intrinsic distribution, and we find
    that this can lead to an inferred distribution that is too
    broad. We use these results and a sample of 25 sources that have
    recent reverberation mapping estimates of AGN black hole masses to
    investigate the effectiveness of using the C IV emission line to
    estimate $M_{BH}$ and to indirectly probe the C IV region
    size--luminosity ($R$--$L$) relationship. A linear regression of
    $\log L_{\lambda} (1549\AA)$ on $\log M_{BH}$ found that $L_{1549}
    \propto M_{BH}^{1.17 \pm 0.22}$. A linear regression also found
    that $M_{BH} \propto L_{1549}^{0.41 \pm 0.07} FWHM_{CIV}^2$,
    implying a C IV $R$--$L$ relationship of the form $R_{CIV} \propto
    L_{1549}^{0.41 \pm 0.07}$. Including the C IV line $FWHM$ resulted
    in a reduction of a factor of $\sim 1/3$ in the error in the
    estimates of $M_{BH}$ over simply using the continuum luminosity,
    statistically justifying its use. We estimated $M_{BH}$ from both
    C IV and H$\beta$ for a sample of 100 sources, including new
    spectra of 29 quasars. We find that the two emission lines give
    consistent estimates if one assumes $R \propto L^{1/2}_{UV}$ for
    both lines.

  \end{abstract}
  
  \keywords{galaxies:\ active\ ---\ line:\ profiles\ ---\ methods:\ data analysis\ ---\ 
    methods:\ statistical\ ---\ quasars:\ emission lines}
  
  \section{INTRODUCTION}

  \label{s-intro}

  It is widely accepted that the extraordinary activity associated
  with quasars involves accretion onto a supermassive black hole
  (SMBH). Furthermore, the evidence that almost all massive galaxies
  host SMBHs has become quite convincing. It has been found that SMBH
  mass is correlated with the host galaxy's bulge luminosity
  \citep[e.g.,][]{korm95,mag98,mclure01,marc03} as well as the stellar
  velocity dispersion
  \citep[e.g.,][]{gebh00a,merr01,mclure02,trem02}. Because luminous
  quasars have been observed to reside in massive early-type galaxies
  \citep{mclure99,kuk01,mcleod01,nolan01,perc01,dunlop03}, this
  implies that the evolution of spheroidal galaxies and quasars is
  intricately tied together
  \citep[e.g.,][]{silk98,haehn00,adams01,merr04,dimatt05}. Therefore,
  understanding the cosmic evolution of SMBHs is an important task of
  modern astronomy.

  Reverberation mapping \citep{bland82,peter93} is often used to
  estimate SMBH mass, $M_{BH}$, in Type 1 active galaxies
  \citep{wand99,kaspi00,peter04}. One of the principal advantages of
  this method is that it does not require high spatial resolution, but
  rather relies on the time lag between the continuum and emission
  line variability. Under the assumption that the broad line region
  (BLR) is in Keplerian motion, the time lag is combined with the line
  width to give an estimate of SMBH mass. This, in principal, makes it
  immediately applicable for both faint and distant quasars. Although
  there are many potential systematic uncertainties in the technique
  \citep{krolik01}, there has been good agreement between SMBH masses
  inferred from reverberation mapping and those inferred from the
  stellar velocity dispersion \citep{gebh00b,ferr01,nelson04,onken04}.

  Unfortunately, reverberation mapping requires long-term intensive
monitoring, which is not practical for large samples. In addition, the
long time scales for variability in bright high redshift sources make
reverberation mapping for these sources unfeasable. Fortunately, a
correlation has been found that relates the BLR size and monochromatic
continuum luminosity \citep[the $R$--$L$ relationship,
][]{kaspi00,kaspi05}, making it possible to estimate virial masses
from single-epoch spectra using the H$\beta$ width \citep{wand99}. In
addition, \citet{wu04} find a relationship between BLR size and the
luminosity of H$\beta$, but \citet{woo02a} do not find any correlation
between SMBH mass and bolometric luminosity. H$\beta$ is redshifted
into the near-infrared at $z \sim 0.9$, making it difficult to observe
from the ground for large samples. \citet{vest02} and \citet{bhmmgii}
have argued for the use of C IV and Mg II, respectively, to estimate
virial masses from single-epoch spectra, allowing quasar SMBH masses
to be estimated at high $z$ from the ground. Many studies have
exploited these results and estimated SMBH masses for large samples of
quasars
\citep[e.g.,][]{bech03,corb03,warn03,vest04,mclure04a,mclure04b}. \citet{diet04}
have found that the estimates based on H$\beta$ and C IV agree well
for high $z$ quasars, whereas Mg II-based estimates are typically a
factor of $\sim 5$ times lower. In constrast to this, \citet{shem04}
and \citet{bask05} argue that C IV does not give as accurate of an
estimate of SMBH mass as H$\beta$.

Previous methods that utilize single-epoch spectra have relied on
empirical linear relationships that estimate the BLR size for a given
source luminosity, and then use this estimate of $R$ in the virial
relationship. Although this will give good estimates for $R$, this
may give less efficient estimates for SMBH mass. The reason for this
is that the luminosity may also be correlated with $M_{BH}$ in a
manner independent of the $R$--$L$ relationship, such as through the
accretion process. By only using the luminosity as a proxy for $R$
when estimating $M_{BH}$, one is ignoring the additional information
of $M_{BH}$ that is contained within $L$. In other words, the standard
broad line estimates are based on the distribution of velocities at a
given luminosity and $M_{BH}$, as folded through the $R$--$L$
correlation and under the assumption of Keplerian motion. However, the
distribution of black hole masses at a given luminosity and line width
is not simply obtained by inverting the distribution of velocities at
a given luminosity and black hole mass, unless $L$ and $M_{BH}$ are
statistically independent.

Recent studies have found evidence for a correlation between
luminosity and black hole mass \citep[e.g.,][]{corb03, netzer03,
peter04}, suggesting that black hole mass and luminosity are not
statistically independent.  However, in contrast to other studies,
\citet{woo02a} have argued that there is no significant correlation
between bolometric luminosity and $M_{BH}$. If a correlation between
$M_{BH}$ and $L$ exists, then we can combine the $M_{BH}$--$L$
correlation with the broad line mass estimates to obtain, on average,
more accurate estimates of $M_{BH}$ for a given luminosity and line
width.

In this work, we outline a formalism that allows one to estimate the
probability distribution of an AGN's black hole mass, given some set
of observables. We focus on the special case of estimating $M_{BH}$
given some monochromatic luminosity, $L_{\lambda}$, and the width of
an emission line. We also search for other parameters of the C IV line
that may contribute additional information of $M_{BH}$, thus
decreasing the uncertainty in the estimated $M_{BH}$. Although it is
possible to include other quasar properties, such as radio loudness,
X-ray loudness, or variability \citep[e.g.,][]{xie05, oneill05,
pessah06}, for simplicity we have chosen to only include parameters
that may be measured from a single spectrum containing C IV. This
allows $M_{BH}$ to be estimated using only one spectrum, and it is
thus not necessary to compile observations from several different
spectral regions.

We use the formalism developed here to justify using the C IV line
width in estimating $M_{BH}$, and attempt to indirectly infer the C IV
$R$--$L$ relationship. We have chosen the C IV line because it is
readily observable from the ground over a wide range in redshift ($1.5
\lesssim z \lesssim 4.5)$, is less effected by blends with iron and
other lines, has shown to give consistent mass estimates with H$\beta$
\citep{vest02}, is commonly employed to estimate SMBH mass, and
archival UV spectra are available for most of the sources with
reverberation-based masses \citep{peter04}. Despite its common usage,
the $R$--$L$ relationship for C IV is mostly unexplored, as there are
only seven data points with reliable C IV reverberation mapping data
\citep{peter05}. Often it is assumed that C IV BLR size has the same
dependence on luminosity as H$\beta$
\citep[e.g.,][]{vest02,netzer03,vest06}. Another possibility is to
assume $R \propto L^{1/2}$, as predicted from simple photoionization
theory or if the BLR size is set by the dust sublimation radius
\citep{netz93}; this was done by \citet{wand99} and \citet{shields03}
for the H$\beta$ line. \citet{peter05} performed a linear regression
using five AGN with a total of seven data points and find $R \propto
L_{UV}^{0.61 \pm 0.05}$, similar to that for the Balmer
lines\footnote{This value appears in an erratum to this
paper}. Unfortunately, this result is almost entirely dependent on the
inclusion of NGC 4395, the least luminous known Seyfert 1 galaxy,
since the data points for the other AGN are clustered around $\lambda
L_{\lambda}(1350\AA) \approx 10^{43.75}\ {\rm ergs\
s^{-1}}$. \citet{vest06} found that using the C IV line to calculate
black hole masses assuming $R \propto L_{UV}^{0.53}$ gave results
consistent with masses obtained by reverberation mapping.

The layout of the paper is as follows. In \S~\ref{s-bhmass} we
describe the general problem of estimating SMBH mass from single-epoch
spectra, and in \S~\ref{s-sample} we describe our two samples. The
first sample is a set of 25 quasars with black holes from
reverberation mapping and archival UV spectra, and the second sample
is a set of 100 quasars for which we have spectra containing both the
H$\beta$ and C IV emission lines. In \S~\ref{s-lineparams} we describe
the method we employ to estimate the emission line parameters. In
\S~\ref{s-varselect} we test if any other UV or C IV parameters
contribute useful information about $M_{BH}$.  In \S~\ref{s-mlrel} we
use the sample with reverberation mapping data to investigate the
$M_{BH}$--$L$ relationship, and in \S~\ref{s-civrl} we test if
including the C IV line $FWHM$ is preferred by this sample and
investigate the nature of the C IV $R$--$L$ relationship.  In
\S~\ref{s-hbeta} we use our larger sample to compare estimates of
$M_{BH}$ obtained from both single-epoch H$\beta$ and C IV. We
summarize our results in \S~\ref{s-summary}.

In this work we adopt the WMAP best-fit cosmological parameters
\citep[$h=0.71, \Omega_m=0.27, \Omega_{\Lambda}=0.73$,][]{wmap}. We
will use the common statistical notation where a point estimate of a
quantity is denoted by placing a $\hat{}$ above it, e.g.,
$\hat{M}_{BH}$ would be an estimate of $M_{BH}$.  We will also
commonly refer to the bias of an estimate, where the bias of
$\hat{M}_{BH}$ is $Bias = E(\hat{M}_{BH}) - M_{BH}$. Here,
$E(\hat{M}_{BH})$ is the expectation value of $\hat{M}_{BH}$. An
unbiased estimate of $M_{BH}$ is one such that $E(\hat{M}_{BH}) =
M_{BH}$.

  \section{ESTIMATING BLACK HOLE VIRIAL MASSES}

  \label{s-bhmass}
  
  If one assumes that the BLR gas is in Keplerian motion, then the
  mass of the central black hole, $M_{BH}$, may be estimated from the
  virial theorem :
  \begin{equation}
    M_{BH} = f \frac{R V^2}{G}.
      \label{virial}
  \end{equation}
  Here $R$ is the distance between the BLR that is emitting a
  particular line and the central continuum source, $V$ is the
  velocity dispersion of the line-emitting gas, and $G$ is the
  gravitational constant.  The velocity dispersion is inferred from
  the width of the line, quantified using either the $FWHM$ or line
  dispersion, $\sigma_*$. The factor $f$ is a scaling factor that
  converts the virial product, $R V^2 / G$, to a mass. Typically, $f$
  has been set to the value appropriate for an isotropic velocity
  field, where $f = 3/4$ if one uses the broad line $FWHM$ to estimate
  $V$ \citep[e.g.,][]{netz90}; however, there may be systematic
  effects that can significantly effect the value of $f$
  \citep{krolik01}. \citet{onken04} used the correlation between
  $M_{BH}$ and stellar velocity dispersion to estimate an average
  scale factor of $\langle f \rangle = 5.5$ when the line dispersion
  is used to estimate $V$. For simplicity, in the rest of this work we
  will assume $f = 1$, so that what is really being estimated is the
  virial product. After estimating the virial product, we can convert
  it to a mass using any adopted value of $f$.

  \subsection{Estimating $M_{BH}$ from Reverberation Mapping}

  \label{s-reverb}

  For the case of reverberation mapping, estimating $M_{BH}$ is
  straightforward. The BLR size, $R$, can be estimated as $c\tau$,
  where $c$ is the speed of light and $\tau$ is either the peak or the
  centroid of the line--continuum cross-correlation function. The
  velocity dispersion of the line-emitting gas, $V$, is estimated from
  the width of the broad emission line as measured in the variable
  part of the spectrum. Measuring $V$ from the variable (RMS) spectrum
  ensures that one is probing the line emission that is actually
  varying, i.e., the BLR gas that is at the distance $c\tau$. One then
  uses $c\tau$ and $V$ in Equation (\ref{virial}) to estimate
  $M_{BH}$. See \citet{peter04} for recent reverberation mapping
  results.

  \subsection{Estimating $M_{BH}$ from Single-Epoch Spectra}

  \label{s-sespec}

  Estimating $M_{BH}$ from a single-epoch spectrum (SES) is a somewhat
  different problem for several reasons. In this section we illustrate
  why the SES case is different, and provide a general formalism for
  calculating the conditional probability distribution of $M_{BH}$,
  given a set of observables.

  In the single-epoch case, one cannot directly observed the BLR size,
  but instead employs a correlation between $R$ and continuum
  luminosity. However, there is considerable scatter in the $R$--$L$
  relationship, which is propogated through when using $L$ instead of
  $R$ to estimate $M_{BH}$. In this case, the conditional probability
  distribution of $M_{BH}$, given the broad line estimate, is broad,
  typically a factor of a few \citep{vest02,kaspi05}. This scatter can
  be reduced by incorporating information about $M_{BH}$ from other
  observables. In the more general case, one may have several
  parameters that contain information about $M_{BH}$, such as emission
  line width, variability, luminosity, etc. For example,
  \citet{merloni03} have found evidence that black hole mass is
  correlated with X-ray and radio luminosity. In this case, one could
  combine the information from the line width, UV luminosity, X-ray
  luminosity, and radio luminosity to obtain an estimate of $M_{BH}$
  that is more accurate than would have be obtained solely from some
  subset of these parameters.

  To be more specific, suppose that one has a set of observables,
  denoted by $X$. Then, given the observables, $X$, the conditional
  probability of $M_{BH}$ given $X$ is given by Bayes' Formula
  \begin{equation}
    p( M_{BH} | X ) = \frac{p( X | M_{BH} ) p( M_{BH} )}{p(X)},
    \label{condprob}
  \end{equation}
  where $p(M_{BH})$ is the intrinsic probability distribution of
  $M_{BH}$ for a sample, and $p(X)$ is the distribution of the
  observables; $p(X)$ is just a normalizing constant and may be
  ignored.

  In this work we are concerned with the distribution of $M_{BH}$
  given $L$, the emission line width, and redshift. For this case,
  Equation (\ref{condprob}) becomes
  \begin{equation}
    p(m|l,v,z) \propto p(v|l,m,z) p(l|m,z) p(m|z).
    \label{mlv}
  \end{equation}
  Here, we are using the notation, $m \equiv \log M_{BH}$, $l \equiv
  \log \lambda L_{\lambda}$, and $v \equiv \log V$.  The first term,
  $p(v|l,m,z)$, is the distribution of line widths at a given $l$,
  $m$, and $z$. The second term, $p(l|m,z)$, is the distribution of
  luminosities at a given $m$ and $z$. The last term, $p(m|z)$, is the
  distribution of $m$ at a given redshift. As we will show later in
  this section, $p(v|l,m,z)$ is obtained by plugging the $R$--$L$
  relationship into the Virial theorem (Eq.[\ref{virial}]). This is
  the standard method of estimating $M_{BH}$ from the broad lines, but
  it implicitly assumes $p(m|l,v,z) \propto p(v|l,m,z)$, and therefore
  that $p(m|z)$ and $p(l|m,z)$ are uniform. Taking $p(m|z)$ and
  $p(l|m,z)$ to be uniform results in a broader distribution of
  $M_{BH}$, given $L, V,$ and $z$, and thus a less efficient, albeit
  still unbiased, estimate of $M_{BH}$. However, by incorporating
  information on both the distribution of luminosities at a given
  black hole mass and redshift, and the distribution of $M_{BH}$ at a
  given redshift, one can obtain a better estimate of $M_{BH}$.

  To estimate a functional form for $p(l|m)$ and $p(v|l,m,z)$, suppose
  we observe some quasar with SMBH mass $M_{BH}$ at a redshift $z$,
  where $m$ is drawn from some probability density $p(m|z)$, $m|z \sim
  p(m|z)$. We assume that the accretion process for this source
  generates a luminosity, $L_{\lambda} \propto M_{BH}^{\alpha_m(z)}$,
  by
  \begin{equation}
    l|m,z = \alpha_0 + \alpha_m(z) m + \epsilon_l(z,m).
    \label{mlum}
  \end{equation}
  Here, $\alpha_0$ is some constant of proportionality, and
  $\epsilon_l(z,m)$ is the random error term representing the scatter
  about this relationship. We will refer to Equation (\ref{mlum}) as
  the $M_{BH}$--$L$ relationship. The stochastic term,
  $\epsilon_l(z,m)$, encompasses variations at a given $M_{BH}$ in the
  accretion rate ($\dot{M}$), radiative efficiency ($\epsilon$),
  source inclination, bolometric correction ($C_{bol}$), etc.; since
  we do not observe these quantities we model them as being
  random. For now, we allow the logarithmic slope, $\alpha_m$, to
  depend on $z$, and the random error to depend on $z$ and
  $M_{BH}$.

  The parameter $\alpha_m$ can be predicted from accretion
  physics. The radiated bolometric luminosity from an accretion flow
  can be written as
  \begin{equation}
    L = ( 1.26 \times 10^{31} ) \epsilon \dot{m} \frac{M_{BH}}{M_{\odot}}\ {\rm W},
    \label{acclum}
  \end{equation}
  where $\dot{m} = \dot{M} / \dot{M}_{edd}$ is the accretion rate
  normalized to Eddington. If one assumes $L_{\lambda} = C_{bol} L$,
  then from Equation (\ref{acclum}) it follows that for this case
  $\alpha_m = 1$ and $\epsilon_l = \log \epsilon + \log \dot{m} + \log
  C_{bol}$. A more careful analysis of the thin disk case suggests
  $\lambda L_{\lambda} \propto (M_{BH} \dot{M})^{\alpha_m}, \alpha_m =
  2/3$, after employing some simplifying approximations
  \citep[e.g.,][]{collin02}.

  Given this luminosity, the BLR distance $R$ is assumed to be set by
  the luminosity according to the $R$--$L$ relationship, $R \propto
  L^{\theta_l}$:
  \begin{equation}
    r|l,z = \theta_0 + \theta_l l + \epsilon_r(z).
    \label{rlum}
  \end{equation}
  Similar to above, $r \equiv \log R$ and $\epsilon_r(z)$ is the
  stochastic component. In Equation (\ref{rlum}) we assume that given
  $L$, $R$ is independent of $M_{BH}$. In addition, we have not
  assumed any redshift dependence for $\theta_l$ because it is likely
  that the form of the $R$--$L$ relationship is independent of $z$
  \citep{vest04}. However, in Equation (\ref{rlum}) we have allowed
  for the possibility of a redshift dependence for $\epsilon_r$ as the
  the intrinsic scatter in the $R$--$L$ relationship may depend on
  $z$.

  Simple photoionization theory predicts that $R \propto
  L_{ion}^{1/2}$, where $L_{ion}$ is the luminosity of the ionizing
  continuum \citep[e.g.,][]{wand99,kaspi00}.  From the definition of
  the ionization parameter, $U$, it follows that
  \begin{equation}
    2 r = \log L_{ion} - \log (4 \pi c) - \log U - \log n_e - \log
    \bar{E},
    \label{ionpar}
  \end{equation}
  where $\bar{E}$ is the average energy of an ionizing photon and
  $n_e$ is the BLR gas density. If we make the simplifying assumptions
  that $r$ is set by Equation (\ref{ionpar}), $L_{\lambda} \propto
  L_{ion}$, and the means of the distributions of $U, n_e,$ and
  $\bar{E}$ are independent of $L_{\lambda}$, then comparison with
  Equation (\ref{rlum}) shows that $\theta_l = 1/2$ and $\epsilon_r =
  -\log U - \log n_e - \log \bar{E}$. If this is not the case, but
  rather the means of the distributions of $U, n_e$, and $\bar{E}$
  have a power-law dependence on $L_{ion}$, then Equation
  (\ref{ionpar}) is still valid, but in general $\theta_l \neq
  1/2$. Either way, in this model the scatter about the $R$--$L$
  relationship is partly the result of variations in $U, n_e,$ and
  $\bar{E}$. Other sources of scatter may include variations in the
  conversion between $L_{\lambda}$ and $L_{ion}$, source inclination,
  and the non-instantaneous response of the BLR to continuum
  variations.

  If $R$ is set by the dust sublimation radius, then we also expect
  $\theta_l = 1/2$ \citep{netz93}, but in this case $R \propto
  L^{1/2}$. A relationship of the form $R \propto L_{\lambda}^{1/2}$
  is consistent with the results of \citet{peter04} for the Balmer
  lines if one uses the UV continuum luminosity.

  Finally, from Equation (\ref{virial}) the observed SES line width
  depends on $R$ and $M_{BH}$ as
  \begin{equation}
    v|r,m = v_0 - \frac{1}{2} r + \frac{1}{2} m + \epsilon_v,
    \label{vrm}
  \end{equation}
  where, $v_0 = \log (\sqrt{G} / f_{SES})$, and $\epsilon_v$ is the
  stochastic term. The term $f_{SES}$ converts the SES line width
  measurement into a velocity dispersion, and $\epsilon_v$ describes
  random deviations of single-epoch $v$ from that for the RMS
  spectrum. \citet{vest02} has found that the single-epoch H$\beta$
  $FWHM$ and the RMS H$\beta$ $FWHM$ are consistent with a scatter of $\sim
  20\%$, if one does not subtract the SES H$\beta$ narrow
  component. When the $FWHM$ is used, $f_{SES} \approx 1/2$. Through
  the stochastic terms $\epsilon_l, \epsilon_r,$ and $\epsilon_v$,
  Equations (\ref{mlum}), (\ref{rlum}), and (\ref{vrm}) define the
  conditional probability densities that describe how $L$ and $V$
  depend on $M_{BH}$, and how $R$ depends on $L$.

  It is useful to examine the special case of Gaussian error terms,
  $\epsilon_l, \epsilon_r,$ and $\epsilon_v$, and Gaussian
  $p(m|z)$. In this case, the distribution of $m$ at a given $l$ and
  $v$ is also normal. If we assume that $\epsilon_l, \epsilon_r,$ and
  $\epsilon_v$ are uncorrelated, have zero mean, and variances
  $\sigma_l^2, \sigma_r^2,$ and $\sigma^2_v$, respectively, then the
  optimal estimate of $m$ is the mean of $p(m|l,v,z)$, $\mu$. If we
  make the further assumption that $\epsilon_l, \epsilon_r,$ and
  $\epsilon_v$ are independent of $m$ and $z$, and that $\alpha_m$ does
  not depend on $z$, then $p(m|l,v,z)$ takes a particularly simple
  form. The mean of $p(m|l,v,z)$ may be calculated from the properties
  of the normal distribution \citep[e.g.,][]{gelman04} as
  \begin{equation}
    \mu = \frac{\hat{m}_{BL} / \sigma^2_{BL} + \hat{m}_{ML} /
      \sigma^2_{ML} + \bar{m}(z) / \sigma^2(z)}{1 / \sigma^2_{BL} + 1
      / \sigma^2_{ML} + 1 / \sigma^2(z)}.
    \label{wmean}
  \end{equation}
  Here, $\hat{m}_{BL} = m_0^{BL} + \theta_l l + 2 v$ is the standard
  broad line mass estimate, found by plugging Equation (\ref{rlum})
  into Equation (\ref{vrm}), $m_0^{BL} = \theta_0 - 2 v_0$,
  $\sigma^2_{BL} = \sigma^2_r + 4 \sigma^2_v$ is the variance in
  $\hat{m}_{BL}$, $\hat{m}_{ML} = (l - \alpha_0) / \alpha_m$ is the
  estimate of $m$ based on the $M_{BH}$--$L$ relationship,
  $\sigma^2_{ML} = \sigma_l^2 / \alpha_m^2$ is the variance in
  $\hat{m}_{ML}$, $\bar{m}(z)$ is the mean $m$ at a given $z$, and
  $\sigma^2(z)$ is the variance in $m$ at a given $z$.
    
  The variance in $\mu$ as an estimate for $m$, $\sigma^2_{\mu}$, is
  \begin{equation}
    \sigma_{\mu}^2 = \left( \frac{1}{\sigma^2_{BL}} +
      \frac{1}{\sigma^2_{ML}} + \frac{1}{\sigma^2(z)} \right)^{-1}.
    \label{wvar}
  \end{equation}
  As is apparent from Equation (\ref{wvar}), the variance in $\mu$ as
  an estimate for $m$ is always less than that for the standard broad
  line estimate, $\hat{m}_{BL}$. In fact, in the limit
  $(\sigma^2_{ML}, \sigma^2(z)) \rightarrow \infty$, $p(m|l)$ and
  $p(m|z)$ supply no information on the black hole mass and $\mu$
  converges to the broad line estimate. However, by combining the
  broad line mass estimate with the information on $M_{BH}$ that is
  contained within the luminosity and redshift we can obtained a
  better estimate of $m$.

  In reality, the distribution of $\dot{m}$ is unlikely to be
  Gaussian, and since $\dot{m}$ is a component of $\epsilon_l$, this
  would violate the assumption that $\epsilon_l$ is Gaussian. Instead,
  the distribution of $\dot{m}$ is likely bimodal
  \citep{ho02,march04,hop06b,cao06} as a result of a transition from a
  radiatively inefficient flow to an efficient one at $\dot{m} \sim
  0.01$ \citep[e.g.,][]{jest05}. However, because of the flux limits
  of modern surveys, most observed broad line quasars will have
  $\dot{m} \gtrsim 0.01$ \citep{mclure04a, vest04}. This results in a
  unimodal, relatively smooth and symmetric distribution of $\dot{m}$
  for observed quasars \citep{hop06a}. It may also be that BLRs do not
  form in sources with $\dot{m} \lesssim 10^{-3}$ \citep{nica03,
  czerny04}, and therefore the distribution of $\dot{m}$ for broad
  line AGN would have a lower limit at $\dot{m} \sim 0.001$. If true,
  then unimodality in the distribution of $\dot{m}$ for broad line AGN
  is ensured.

  The stochastic scatter about the $M_{BH}$--$L$ relationship,
  $\epsilon_l$, is the sum of random deviations in $\dot{m}$,
  bolometric correction, radiative efficiency, etc. We assume that the
  distributions of most, if not all, of the constituent components of
  $\epsilon_l$ are not too different from a Gaussian distribution,
  i.e., unimodal, smooth, and fairly symmetric. Then, by the central
  limit theorem, the distribution of the sum of these components,
  $\epsilon_l$, will tend toward a normal density. Therefore, without
  any evidence to the contrary, the normal density should provide an
  accurate approximation to the true form of $p(l|m)$. A similar
  argument may be used to justify the assumption of normality for
  $\epsilon_r$ and $\epsilon_v$.

  To illustrate the improvement that Equation (\ref{wmean}) offers
  over the broad line mass estimate, we simulate values of $m$, $l$,
  and $v$. The simulations were performed as follows. First, we
  simulated values of $m$ from a smoothly-connected double power-law,
  with a mean of $\bar{m} \approx 7.87$ and a dispersion of $\sigma
  \approx 0.45$ dex. Values of $l$ were then generated according to
  Equation (\ref{mlum}), with $\alpha_0 = 37, \alpha_m = 1$. The
  gaussian scatter about this relationship had a dispersion of 0.7
  dex. Then, we generated values of $v$ as $v = 11 - l / 4 + m / 2 +
  \epsilon$, where $\epsilon$ was a gaussian random variable with
  dispersion 0.2 dex. This form for $v$ assumes $R \propto L^{1/2}$,
  and corresponds to a intrinsic scatter in the broad line mass
  estimates of 0.4 dex. The parameters for simulating $l$ and $v$ were
  chosen to be similar to the results found in
  \S~\ref{s-testing}. Finally, we calculated broad line mass estimates
  from the simulated luminosities and line widths, $\hat{m}_{BL} = -22
  + l / 2 + 2v$, and mass estimates $\mu$ according to Equation
  (\ref{wmean}).

\begin{figure}
    \begin{center}
      \includegraphics[scale=0.4,angle=90]{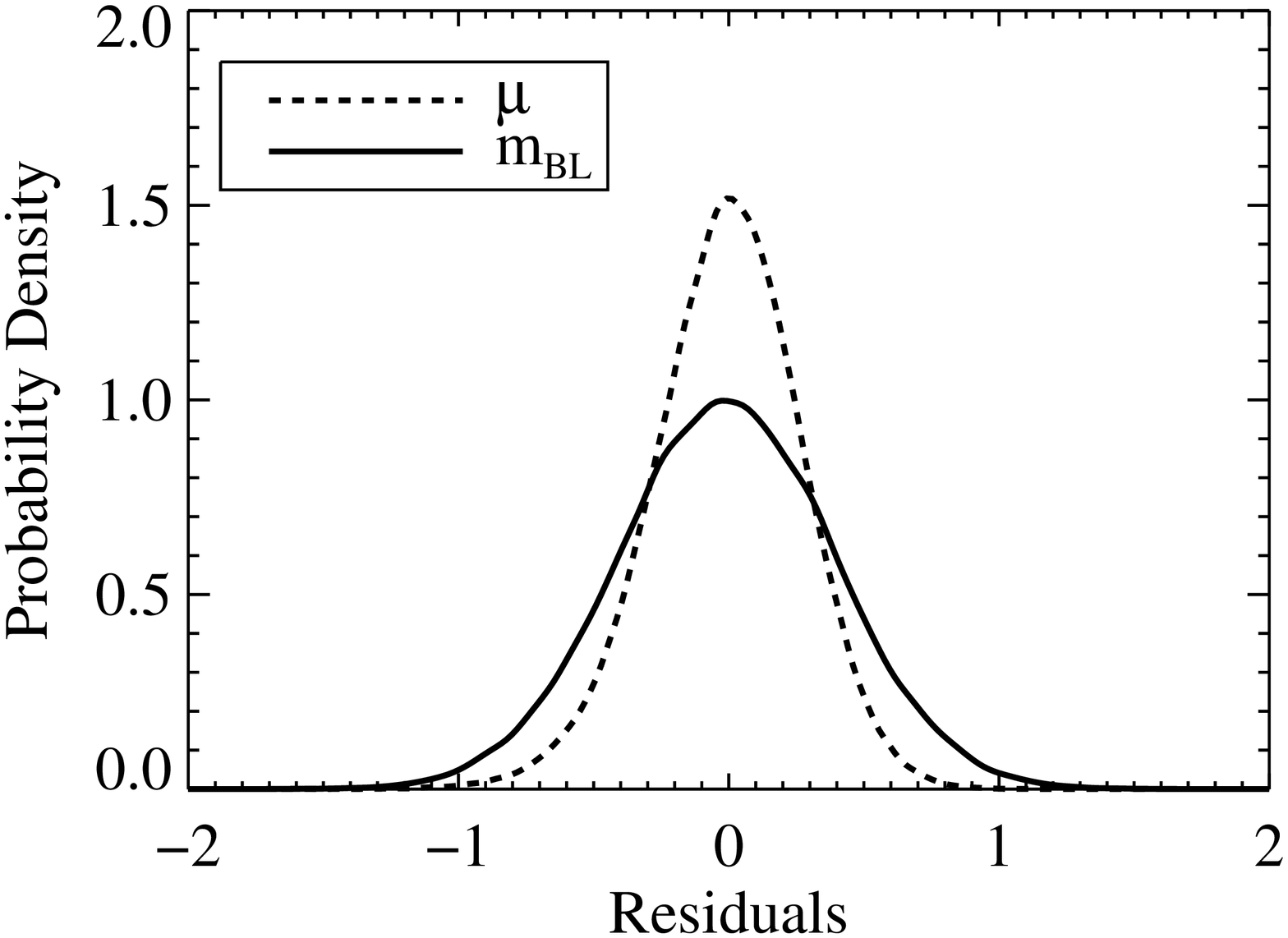}
      \includegraphics[scale=0.4,angle=90]{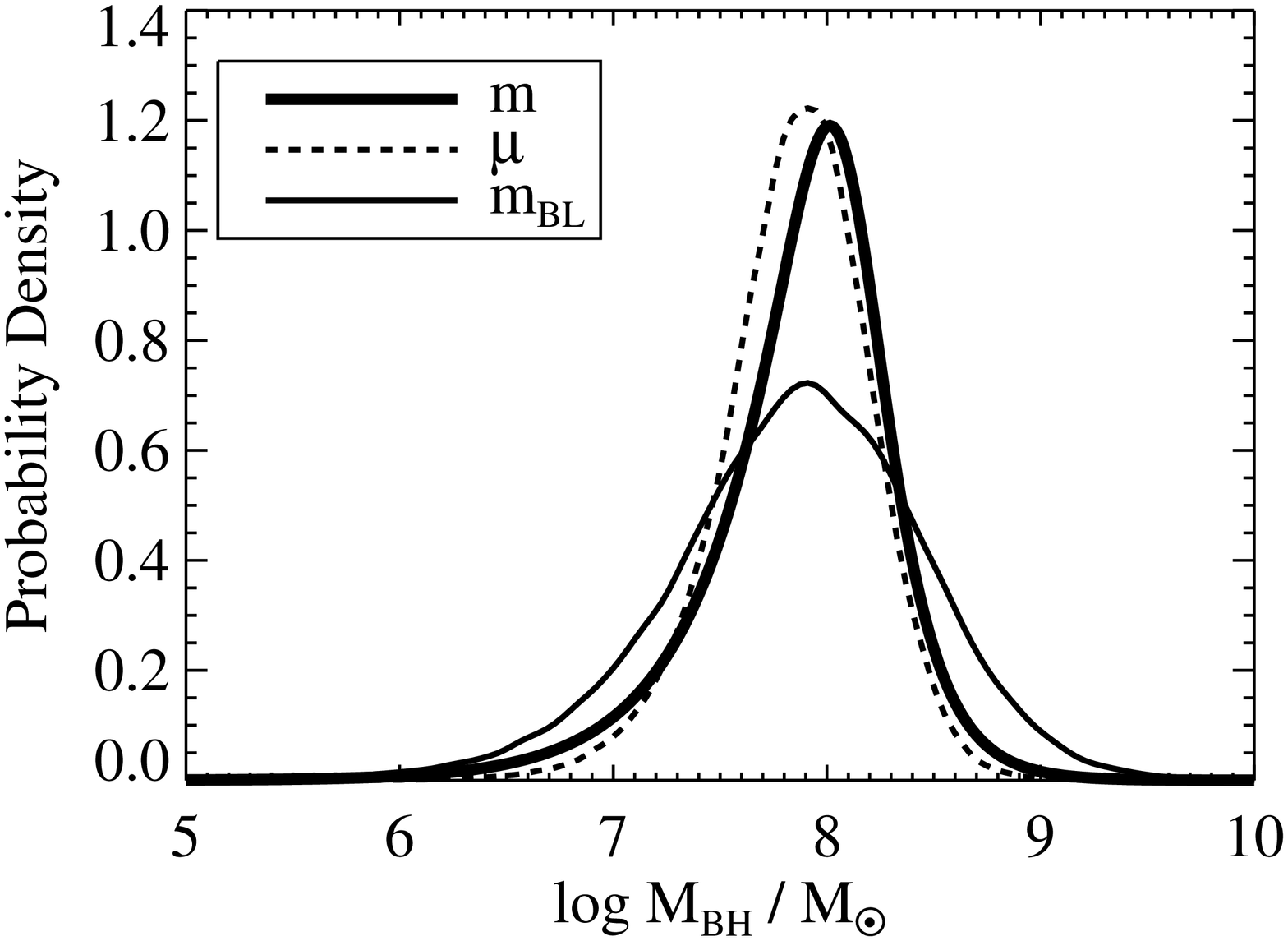}
      \caption{Results of the simulations described in
      \S~\ref{s-sespec}, illustrating the effectiveness of the mass
      estimate, $\mu$, that combines the broad line mass estimates
      with the distribution of luminosities at a given black hole mass
      and the intrinsic distribution of $M_{BH}$. The upper panel shows
      the distribution of the residuals when using $\mu$ (dashed line)
      and the broad line mass estimate, $\hat{m}_{BL}$ (solid
      line). The bottom panel compares the intrinsic distribution of
      $m$ (thick solid line) with that inferred from the distribution
      of $\mu$ (dashed line) and $\hat{m}_{BL}$ (thin solid
      line).\label{f-sim}}
    \end{center}
  \end{figure}

  The results are shown in Figure \ref{f-sim}. As can be seen from the
  distribution of residuals, the mass estimates that combine all
  available information on the black hole mass, $\mu$, are more
  accurate on average than the broad line estimates,
  $\hat{m}_{BL}$. In addition, the distribution of $\mu$ provides a
  more accurate estimate of the true distribution of $m$ than does
  $\hat{m}_{BL}$, with the distribution of $m$ inferred from
  $\hat{m}_{BL}$ being too broad. It is interesting to note that both
  of these results are in spite of the fact that the intrinsic
  distribution of $m$ is not Gaussian, which was assumed when deriving
  Equation (\ref{wmean}). This suggests that if the intrinsic
  distribution of $m$ for a sample is not too different from a normal
  density, Equation (\ref{wmean}) will still give more efficient
  estimates than the broad line estimates. This is reasonable,
  considering that Equation (\ref{wmean}) `shrinks' the black hole
  mass estimates towards the sample mean by an amount inversely
  proportional to the intrinsic variance in $m$ of a sample.

  \subsection{Cautions for Using Quantities Calculated from the SES Mass Estimates}

  \label{s-corrprob}

  The intrinsic uncertainty in $M_{BH}$ inferred from $L_{\lambda}$
  and the line width may be thought of as the `measurement' error in
  $M_{BH}$. This intrinsic uncertainty can cause problems when using
  the black hole mass estimates, $\hat{m}$, to calculate additional
  quantities. In particular, quantities based on the square of the
  estimated $\hat{m}$, such as correlation coefficients and linear
  regressions, can be significantly effected \citep[e.g., see][]{bces,
  fox97}.

  Suppose we are interested in calculating the correlation between
  $M_{BH}$ and some other parameter $X$. To estimate this correlation,
  we would obtain a sample of quasars with black hole masses estimated
  from some assumed form of $p(m|l,v,z)$, $\hat{m}$. Typically, these
  are the standard broad line estimates. We then calculate the
  correlation coefficient between $X$ and $\hat{m}$. However, because
  we do not have the actual black hole masses for our sample, but
  instead obtained estimates from the continuum luminosities and
  widths of the broad lines, this is not the \emph{true} correlation
  coefficient between $X$ and $m$. Because our estimated black hole
  masses have been `measured' with error, this broadens the observed
  distribution of black hole masses, and thus biases the observed
  correlation coefficient towards zero. Therefore, the correlation
  coefficient obtained from the estimated black hole masses will be,
  on average, \emph{less} in magnitude than the correlation
  coefficient that would have been obtained using the actual black
  hole masses.

  To prove this point, we note that the logarithmic black hole mass
  estimates are related to the actual black hole masses as $\hat{m} =
  m + \epsilon_m$, where $\epsilon_m$ is the random error between
  $\hat{m}$ and $m$. The linear correlation between the parameter of
  interest $X$ and $m$ is $\rho = Covar(X,m) / [Var(X) Var(m)]^{1/2}$,
  where $Covar$ and $Var$ are the sample covariance and variance,
  respectively. Since we don't actually observe $m$, we can't
  calculate the true correlation coefficent, but instead we calculate
  the correlation between $X$ and $\hat{m}$, $\hat{\rho}$. While the
  covariance between $X$ and $m$ is unaffected by using $\hat{m}$
  instead of $m$, the sample variance of $\hat{m}$ is $Var(\hat{m}) =
  Var(m) + \sigma_{\hat{m}}^2$, where $\sigma_{\hat{m}}^2$ is the
  intrinsic uncertainty in $\hat{m}$ as an estimate for $m$,
  $\sigma^2_{\hat{m}} = E(\epsilon_m^2)$. The observed correlation is
  then given by
  \begin{equation}
    \hat{\rho} = \left[ \frac{Var(m)}{Var(m) + \sigma^2_{\hat{m}}}.
      \right]^{1/2} \rho
    \label{obscorr}
  \end{equation}
  Therefore, correlation coefficients calculated using the estimated
  black hole masses will be reduced in magnitude from the true
  correlation by a factor of $\left [ 1 + \sigma^2_{\hat{m}} / Var(m)
  \right ]^{-1/2}$. For broad line mass estimates based on H$\beta$,
  \citet{vest06} find $\sigma_{\hat{m}} = 0.43$ dex. In this work we
  find that $\sigma_{\hat{m}} \approx 0.40$ dex for C IV-based broad
  line estimates . For a sample with an intrinsic dispersion in $m$ of
  0.75 dex, these values of $\sigma_{\hat{m}}$ correspond to a
  decrease in the magnitude of any observed correlation by $\approx
  12\%$. If the sample has an intrinsic dispersion of 0.4 dex, similar
  to the intrinsic uncertainties in the broad line estimates, then the
  magnitude of the observed correlation coefficient is reduced by
  $\approx 30\%$. These effects is not negligible, and can be more
  serious for linear regression \citep[e.g.,][]{fox97}. In light of
  these issues, care must be taken when calculating quantities from
  black hole mass estimates based on single-epoch spectra.

  Another problem arises when one is using a flux limited sample to
  estimate the intrinsic distribution of the black hole mass, i.e.,
  the active black hole mass function \citep{wang06}, based on broad
  line estimates. There is currently significant interest in this
  problem, as the active black hole mass function is an important tool
  in understanding SMBH formation and evolution. Unfortunately, the
  limiting flux of a survey causes incompleteness in black hole mass,
  the degree of which depends on the distribution of $l$ at a given
  $m$ and $z$, $p(l|m,z)$. In addition, because the broad line mass
  estimates are measured with error, the distribution of $m$ inferred
  from the broad lines can be significantly broader than the intrinsic
  distribution (cf., Fig.\ref{f-sim}). Because of these issues,
  estimates of the active black hole mass function obtained from broad
  line estimates should be interpreted with caution.

  In order to estimate the degree of incompleteness in $m$, it is
  necessary to re-express the survey selection function as a function
  of $m$.  Following \citet{gelman04}, we introduce an indicator
  variable, $I$, denoting whether a source is included in the survey,
  where $I = 1$ if a is source included and $I = 0$ if a source is
  missed. Then the selection function of the survey is the probability
  that a source is included in the survey for a given luminosity and
  redshift, $p(I=1|l,z)$. The selection function for black hole mass
  is then
  \begin{equation}
    p(I=1|m,z) = \int_{-\infty}^{\infty} p(I=1|l,z) p(l|m,z) dl.
    \label{mselfunc}
  \end{equation}
  As can be seen, estimating the completeness in $m$ for a survey
  depends on the form of $p(l|m,z)$. Therefore, it is important to
  understand $p(l|m,z)$, even if such an understanding does not result
  in significantly better estimates of $m$.

  \section{THE SAMPLE}

  \label{s-sample}

  We employ two samples in our analysis. The first sample consists of
  a set of 25 low-$z$ sources that have reverberation mapping data
  from \citet{peter04}. We use this sample to investigate the
  $M_{BH}$--$L$ relationship and the C IV $R$--$L$ relationship, and
  to justify using the C IV line to estimate AGN black hole
  masses. The second sample is a set of 100 quasars for which we have
  seperate spectra containing the H$\beta$ and C IV emission lines. We
  use this sample to compare the broad line estimates of $M_{BH}$
  obtained from the two lines.

  \citet{peter04} calculated virial products for 35 AGNs based on the
  reverberation mapping method. Of those 35 sources, we selected ones
  with archival UV spectra. We did not include those sources which
  \citet{peter04} listed as having unreliable virial products (PG
  0844+349, PG 1211+143, PG 1229+204, and NGC 4593). We also did not
  include NGC 3227 or NGC 4151, as the C IV line for these sources had
  significant absorption.  In addition, we removed NGC 4051 from the
  analysis because this source is an outlier in the BLR R-L
  relationship \citep{vest02, kaspi05}, and thus probably does not
  follow the linear relationship assumed in Equation (\ref{rlum}).
  Our sample consists of 25 AGN: 14 sources with HST FOS spectra, 9
  sources with IUE spectra, 1 source with HST GHRS spectra, and 1
  source with HST STIS spectra. Of the 14 sources with FOS spectra, 13
  were taken from \citet{bech02}, and the other (NGC 5548) from
  \citet{evans04}. For archival sources with more than one
  observation, we took the source with the longest exposure time. All
  spectra are single-epoch, except for NGC 5548, which is averaged
  over approximately a month of HST observations. The sample is
  summarized in Table \ref{t-data}.

  The luminosities were calculated from the predicted continuum flux
  at $1549 \AA$, assuming a power law continuum (see
  \S~\ref{s-civ_extract}). We corrected luminosities for galactic
  absorption using the $E(B-V)$ values taken from \citet{schlegel}, as
  listed in the NASA/IPAC Extragalactic Database (NED), and the
  extinction curve of \citet{ccm89}, assuming a value of $A_V / E(B-V)
  = 3.1$. We did not do this for the \citet{bech02} sources as they
  have already been corrected for galactic absorption. The C IV line
  widths for the IUE sources were corrected by subtracting an assumed
  instrumental resolution of $1000 \ {\rm km\ s}^{-1}$ in quadrature
  from the measured line widths; the resolution for the other
  instruments is negligible compared to the emission line widths, so
  no correction was performed.

  We compiled UV and optical spectra for a sample of 100 sources for
  the purpose of comparing the C IV-based estimates derived here with
  the estimates based on H$\beta$ and the empirical $R$--$L$
  relationship. Of these sources, 89 have $z < 0.8$, 6 have $z \sim
  2.3$, and 2 have $z \sim 3.3$. The UV spectra for the $z < 0.8$
  sources are FOS spectra from \citet{bech02}, the optical spectra for
  51 of these sources are from \citet{marz03}, and the optical spectra
  for 9 of these sources are from the SDSS DR2 \citep{dr2}. Optical
  spectra for the 29 remaining $z < 0.8$ sources were obtained by
  us. Twenty-seven of the sources were observed at the Steward
  Observatory 2.3m Bok Telescope on Kitt Peak using the 600 lines
  mm$^{-1}$ grating of the B\&C Spectrograph; these sources had
  moderate spectral resolution ($\sim 5\ \AA$). The other two sources
  were observed at the Magellan Baade Telescope using the Inamori
  Magellan Areal Camera and Spectrograph (IMACS); long-slit spectra
  were obtained for these sources using a slit width of 0.9'' in long
  camera mode. We used the 600 lines mm$^{-1}$ grating for PKS
  1451-375 and the 300 lines mm$^{-1}$ grating for PKS 2352-342,
  giving spectral resolutions of $\sim 2\ \AA$ and $\sim 5\ \AA$,
  respectively. The log of new spectra is displayed in Table
  \ref{t-newobs}, and they are shown in Figure 4. The spectra were
  reduced using the standard IRAF routines.

  Rest-frame UV spectra for the eight sources with $z > 2$ are from
  \citet{scott00} and the SDSS. The continuum luminosities and
  H$\beta$ $FWHM$ for the $z \sim 2.3$ sources were taken from
  \citet{mcint99} and corrected to our adopted cosmology. The
  continuum fluxes for the two $z \sim 3.3$ sources were measured
  directly off the published spectra of \citet{diet02}, and the values
  of H$\beta$ $FWHM$ for them are from \citet{diet04}.

  \section{LINE PROFILE PARAMETERS}

  \label{s-lineparams}
  
  \subsection{Extracting The Line Profile}
  
  \label{s-civ_extract}
  
  In order to extract the C IV emission line, it is necessary to
  subtract the continuum, Fe emission, and the He II $\lambda 1640$
  and O III] $\lambda 1665$ emission lines. To remove the continuum
  and iron emission, we use a variation of the method outlined in
  \citet{bg92}. We model the continuum as a power law of the form
  $f_{\nu} \propto \nu^{\alpha}$. The Fe emission was modeled as a
  scaled and broadened form of the Fe emission template extracted from
  I Zw I by \citet{uvfe}. In constrast to most previous studies, we
  simultaneously fit the continuum and Fe emission parameters using
  the Levenberg-Marquardt method for nonlinear $\chi^2$-minimization;
  this is similar to the method used by \citet{mcint99}. Performing
  the fits in this manner has the advantage of providing an estimate
  of the uncertainties in these parameters, given by the inverse of
  the curvature matrix of the $\chi^2$ space. The set of possible
  windows used to fit the continuum and Fe emission are shown in Table
  \ref{t-contwin}. The actual continuum windows used to fit the
  continuum and Fe emission for any particular source depended on that
  source's available spectral range.
  
  Having obtained an estimate of the continuum and Fe emission, we
  subracted these components from each spectrum.  We then extracted
  the region within $-2 \times 10^4 \ {\rm km \ s}^{-1}$ and $3 \times
  10^4 \ {\rm km \ s}^{-1}$ of $1549 \AA$. Here, and throughout this
  work, we will use the convention that negative velocities are
  blueward of a given wavelength. Narrow absorption lines were removed
  and interpolated over. We then removed the He II $\lambda 1640$ and
  O III] $\lambda 1665$ emission lines from the wings of the C IV
  profile. This was done by modelling the C IV, He II, and O III]
  lines as a sum of Gaussians. In general, C IV was modelled as a sum
  of three Gaussians, He II two Gaussians, and O III] a single
  Gaussian, however this varied from source to source.  The C IV
  extraction was done interactively in order to ensure accuracy of the
  fits. After obtaining estimates of the He II and O III] profiles, we
  subtracted these components. We did not fit the N IV] $\lambda 1486$
  emission line as this line is typically weak and lost in the C IV
  wings.
	      
  Extraction of the H$\beta$ profile was done in a similar manner. For
  the optical Fe emission we used the I Zw I template of
  \citet{optfe}. After subtracting the continuum and Fe emission, we
  extracted the region within $\pm 2 \times 10^4 \ {\rm km \ s}^{-1}$
  of $4861 \AA$. The H$\beta$ profile was modeled as a sum of 2--3
  Gaussians. The [O III] $\lambda 4959 \AA$ and [O III] $\lambda 5007
  \AA$ lines were modeled as a sum of 1--2 Gaussians, depending on the
  signal-to-noise of the lines. A sum of two Gaussians was used for
  the higher $S/N$ lines because the [O III] lines are not exactly
  Gaussian, and not because the seperate Gaussians are considered to
  be physically distinct components.  The widths of the narrow
  Gaussians for [O III] were fixed to be equal. The [O III] lines and
  the narrow H$\beta$ line were then subtracted to extract the
  H$\beta$ line. As with the C IV profile, the entire extraction was
  done interactively.
  
  \subsection{Estimating The Line Profile Parameters}
	      
  \label{s-civ_est}

   We measured the C IV line shift, $FWHM$, $EW$, and the first and
   second line moments. We define the line shift, $\Delta v$, as the
   location of the line peak relative to $1549 \AA$, in km
   s$^{-1}$. The location of $1549 \AA$ is determined from the
   redshifts listed on NED. We have checked the references for the
   redshifts to investigate how $z$ was estimated, but not all of the
   references report this. For those that did specify, the redshifts
   were determined from the narrow emission lines (e.g., [O III]
   $\lambda 5007$.).

   IUE observations were done using a very large aperture, so the
   values of $\Delta v$ for the IUE sources may be biased. However, there
   is no noticeable difference between $\Delta v$ estimated from the IUE
   spectra, and those estimated from the HST spectra. Furthermore, the
   $\Delta v$ parameter does not enter into our final analysis, and so
   even if the IUE $\Delta v$ parameters are significantly biased our
   conclusions remain unchanged.

   The first moment of the C IV line is the centroid, $\mu_{CIV}$, and
   the square root of the second central moment is the line
   dispersion, $\sigma_*$. The zeroth line moment is the line
   flux. The line moments are calculated as
   \begin{eqnarray}
     F & = & \sum_{i=1}^n y_i \delta \lambda \label{flux} \\ \mu_{CIV} & = &
     \sum_{i=1}^n x_i y_i \left / \sum_{i=1}^n y_i \label{cent}
     \right. \\ \sigma_*^2 & = & \sum_{i=1}^n x_i^2 y_i \left /
     \sum_{i=1}^n y_i - \mu_{CIV}^2 \right. . \label{rms}
   \end{eqnarray}
   Here, $F$ is the line flux, $\delta \lambda$ is the spacing between
   subsequent wavelengths, $y$ is the observed spectral flux density,
   and $x$ is the velocity relative to $1549\AA$. In practice we do
   not perform the sums over all the $n$ data points, but rather only
   over those data points with $\hat{f}_i \geq 0.05 (\max \hat{f})$,
   and assuming that the profile is monotonically decreasing blueward
   and redward of the peak. Here, $\hat{f}$ denotes the best-fit line
   profile, found from modelling the emission lines as a sum of
   Gaussians. This allows us to define the extent of the line
   profile. Although this gives us biased measurements of the line
   moments, it keeps these measurements stable. In particular,
   $\sigma_*$ can be very sensitive to the profile wings, which have
   the highest uncertainty. By truncating the line profile we keep the
   estimate stable and less sensitive to errors in the Fe and
   continuum subtraction, as well as errors in the He II $\lambda
   1640$ and O III] $\lambda 1665$ subtraction.

   The line moments are calculated using the observed line profile,
   $y$, and not the best-fit to the spectral flux densities,
   $\hat{f}$. We do this because the line moments are relatively
   insensitive to a lack of smoothness. To be specific, consider a
   line profile, $f(x)$, and its Fourier transform,
   $\tilde{f}(k)$. For simplicity, we consider the continuous case
   here. The $j^{th}$ unnormalized line moment, $\mu_j$, may be
   written in terms of the Fourier transform of the line profile:
   \begin{equation}
     \mu_j = \int_{-\infty}^{\infty} x^j f(x) dx =
     \frac{\tilde{f}^{(j)} (0)}{(-2 \pi i)^j},
     \label{fftmom}
     \end{equation}
   where $\tilde{f}^{(j)} (k)$ is the $j^{th}$ derivative of
   $\tilde{f}(k)$ and $i=\sqrt{-1}$. One can see from Equation
   (\ref{fftmom}) that the line moments only depend on $\tilde{f}(k)$
   near $k=0$, and are thus insensitive to the high frequency behavior
   of $f(x)$. A generic smoothing operator, such as a Gaussian fit,
   will shrink the high $k$ components of $\tilde{f}(k)$ more than the
   low $k$, as it is generally the case that the high $k$ components
   have lower signal-to-noise. However, because the line moments do
   not depend on the high $k$ components, nothing is gained by
   enforcing smoothness.  Because of this we just use the observed C
   IV profile, $y$, as it is an unbiased estimate of the true profile,
   $f$, whereas the best-fit estimate, $\hat{f}$, is a biased
   estimate.

   There has been some discussion in the literature over whether
   $FWHM$ or $\sigma_*$ is the better width to use in estimating $m$
   \citep{from00,peter04}. \citet{from00} suggested using the line
   dispersion, $\sigma_*$, arguing that it provides a better estimate
   of the velocity dispersion for an arbitrary line profile. Other
   advantages of $\sigma_*$ include its insensitivity to noise and
   narrow absorption lines. However, $\sigma_*$ can be significantly
   biased due to its sensitivity to blending with other lines in the
   line wings, and to the removal of continuum and iron
   emission. These facts can be understood in light of Equation
   (\ref{fftmom}), which shows that the second moment of the line
   depends on the low-$k$ behavior. Because of this, $\sigma_*$ is
   relatively unaffected by information on small scales, such as noise
   and narrow absorption lines, but is significantly affected by
   information on large scales, such as line blending, truncation of
   the line profile, and continuum placement. The $FWHM$, on the other
   hand, is insensitive to line blends and errors in the continuum and
   iron subtraction. In addition, $FWHM$ is easily measured, however
   it probably provides a poor estimate of the velocity dispersion for
   irregularly shaped lines profiles. \citet{peter04} compared the
   strengths and weaknesses of these two measurements and concluded
   that $\sigma_*$ was the better parameter when measured in the RMS
   spectrum.

   In order to choose the better SES estimate of the BLR velocity
   dispersion, we calculate the partial correlation between $m$ and
   $FWHM$ and $\sigma_*$, respectively. The partial correlation
   coefficient describes the correlation between $m$ and line width at
   a given luminosity; the line width with the higher partial
   correlation should give a better estimate of $m$. The partial
   linear correlation between $\log FWHM$ and $m$ is 0.36, while the
   partial linear correlation between $\log \sigma_*$ and $m$ is
   0.31. Because the $FWHM$ has a moderately higher partial
   correlation, and because the $FWHM$ is not as affected by errors in
   the line deblending, continuum placement, etc., we use the $FWHM$
   as an estimate of the velocity dispersion throughout the rest of
   this analysis.

   We only measure the $FWHM$ of the H$\beta$ emission line. This is
   because we are only concerned with getting an estimate of $M_{BH}$
   from single-epoch H$\beta$ based on the virial theorem for
   comparison with our C IV-based estimates, so the only line
   parameter of interest is the H$\beta$ width.

   The standard errors on the line parameter measurments are estimated
   using the bootstrap \citep{efron79}. In this method, we take our
   best-fit line profile spectral flux densities, $\hat{f}$, and
   generate $n_{\rm boot}=128$ simulated observed line profiles by
   adding Gaussian noise to $\hat{f}$. We then estimate the line
   parameters of the simulated line profiles and calculate the
   variance in these parameters over the bootstrap samples.

   In Figure \ref{f-mvx} we plot $\log M_{BH}$ against $\log \lambda
   L_{\lambda} (1549\ \AA)$, $\log FWHM, \log \sigma_*, \log EW,
   \Delta v, \mu_{CIV},$ and continuum spectral slope. We report our
   measurements in Table \ref{t-params}. We have compared our
   measurements with \citet{wang96}, \citet{bech02}, \citet{bask04},
   and \citet{kuras04}, and find them to be consistent after
   accounting for the different procedures used to measure these
   quantities. For clarity, we have removed three outliers in $\alpha$
   from the plot of $m$ against $\alpha$. These sources were 3C 390.3,
   NGC 3516, and NGC 7469. 3C 390.3 is a broad-line radio galaxy with
   highly variable Balmer lines and double-peaked H$\alpha$
   \citep{corb98} and H$\beta$ \citep{oster76} profiles; the C IV
   emission line also exhibits a double-peaked profile. The other two
   sources represent some of the faintest sources in our sample, and
   their spectra may have a contribution from their host galaxies;
   however, we notice nothing unusual about their spectra, save for
   their unusually soft values of $\alpha$.

  \section{SELECTING THE IMPORTANT PARAMETERS}

  \label{s-varselect}

  It is useful to investigate whether any of the additional parameters
  that we measured for C IV are significantly correlated with $m$, and
  thus contribute information about $m$ in addition to that in $l$ and
  $FWHM$. We used our sample of 25 sources with $M_{BH}$ from
  reverberation mapping to test if including $\log EW, \alpha,
  \mu_{CIV},$ or $\Delta v$ resulted in more accurate estimates of
  $M_{BH}$. However, we did not find any evidence to warrant the
  inclusion of these parameters.

  In order to assess whether the data support including any additional
  parameters, we express $m$ as a linear combination of all possible
  combinations of $\log \lambda L_{\lambda}, \log FWHM, \log EW,
  \alpha, \mu_{CIV},$ and $\Delta v$, a total of $2^6 = 64$
  subsets. The regression coefficients are estimated via
  least-squares, and the result is a set of 64 linear regressions.  In
  order to assess the relative merits of each of the regressions, and
  thus each of the respective parameters, we employ the Bayesian
  Information Criterion \citep[$BIC$][]{bic}. Using the $BIC$ allows
  us to undertake a Bayesian comparison of the models without actually
  carrying out the full Bayesian prescription, considerably
  simplifying things. The $BIC$ is easily obtained from the
  log-likelihood of the data as
  \begin{equation}
    BIC = 2 \ell(\hat{\psi}) + d \ln n.
    \label{e-bic}
  \end{equation}
  Here, $\ell(\hat{\psi})$ is the log-likelihood of the data evaluated
  at the maximum-likelihood estimate, $\psi$ denotes the regression
  parameters, $d$ is the number of parameters in the regression, and
  $n$ is the number of data points.

  The only parameter significantly correlated with $m$ is $l$, with a
  posterior probability of $p_l = 0.975$. The data are ambiguous as to
  whether $FWHM$ is related to $M_{BH}$ ($p_{FWHM} = 0.562$); however,
  when using the $BIC$ we did not compare with regressions that assume
  $M_{BH} \propto FWHM_{CIV}^2$, and we perform a more careful
  analysis in \S~\ref{s-civrl}. In addition, the data give weak
  evidence that the remaining parameters are unrelated to $m$. The
  posterior probabilities that $m$ is correlated with these parameters
  are estimated to be 0.270, 0.210, 0.198, and 0.186 for $\log EW$,
  $\mu_{CIV}$, $\Delta v$, and $\alpha$, respectively.

  \section{REGRESSION ANALYSIS}

  \label{s-testing}
  
  Now that we have ruled out including any additional parameters for
  estimating $M_{BH}$ from SES, we proceed to investigate the
  $M_{BH}$--$L$ relationship and the C IV $R$--$L$
  relationship. Because $p(l|m)$ and $p(v|l,m)$ are statistically
  independent in their parameters, we can analyze each one
  seperately. Throughout this section we will be using our sample of
  25 sources with black hole mass measurements from reverberation
  mapping. We begin by investigating the $M_{BH}$--$L$ relationship.

  \subsection{The $M_{BH}$--$L$ Relationship}

  \label{s-mlrel}

  We fit a linear relationship of the form $l = \alpha_0 + \alpha_m
  m$, assuming that the scatter about this relationship is independent
  of $m$ and $z$. Because we are fitting the distribution of $l$ at a
  given $m$, we use the BCES($Y|X$) \citep{bces} regression. The BCES
  methods take into account measurement errors in both coordinates by
  correcting their moments; however, the measurement errors are small
  compared to the variance in both $m$ and $l$, so the correction is
  small. Based on the regression, we find
  \begin{equation}
    l = 35.72( \pm 1.67) + 1.17( \pm 0.22) m .
    \label{mlumemp}
  \end{equation}
  Here, $l = \log \lambda L_{\lambda} (1549\AA)$. The empirical value
  of $\hat{\alpha}_m = 1.17 \pm 0.22$ is consistent with $L \propto
  M_{BH}$, if one assumes that $L_{\lambda} \propto L$. The intrinsic
  scatter in Equation (\ref{mlumemp}) is estimated to be
  $\hat{\sigma}_l = 0.61$ dex. The residuals and their cumulative
  distribution function (CDF) are shown in Figure \ref{f-mlumresid}. A
  Kolmogorov-Smirnov test found that the regression residuals are not
  significantly different from a normal distribution, implying that
  $p(l|m)$ is normal with mean given by Equation (\ref{mlumemp}) and
  standard deviation $\sigma_l \approx 0.61$ dex.

  The value of $\hat{\alpha}_m$ found here is consistent with $M_{BH}
  \propto L^{0.9 \pm 0.15}$ found by \citet{netzer03} and $M_{BH}
  \propto L^{0.97 \pm 0.16}$ found by \citet{corb03}. \citet{peter04}
  used the 35 AGN with reverberation mapping data to estimate the
  $M_{BH}$--$L$ relationship for $L_{\lambda}$ at $5100\AA$. Using the
  BCES bisector regression, they find $M_{BH} \propto L_{5100}^{0.79
  \pm 0.09}$. This is shallower than our result of $M_{BH} \propto
  L_{1549}^{1.17 \pm 0.22}$, although the two logarithmic slopes are
  consistent within the errors. The difference in the two values most
  likely results from the different regressions used. We used the
  BCES($Y|X$) because we are modelling $p(l|m)$, where as the bisector
  slope gives the regression that bisects the distribution of $l$ at a
  given $m$, and of $m$ at a given $l$.

  \citet{woo02a} did not find any evidence for a correlation between
  $M_{BH}$ and the bolometric luminosity. They used both a sample of
  reverberation-mapped AGNs and Seyfert galaxies with black hole
  masses derived from the stellar velocity dispersion. In neither case
  did they find evidence for a correlation, in constrast to the
  results found here and by others. Unfortunately, \citet{woo02a} did
  not perform a regression analysis or report a correlation
  coefficient, so it is difficult to do a quantitative comparison of
  their results with ours.

  We have found here that $\hat{\alpha}_m = 1.17 \pm 0.22$ for the
  \citet{peter04} sample. However, these sources are all at low
  redshift and have $\dot{m} = 0.01$--$1$ \citep{woo02a,peter04}, and
  thus this value of $\alpha_m$ may not be valid for sources with
  $\dot{m} \lesssim 0.01$ and $z \gtrsim 0.2$. Furthermore, most high
  $z$ broad line quasars have luminosities and $M_{BH}$ greater than
  that of the \citet{peter04} sample, and it is possible that
  $\alpha_m \neq 1$ outside of the reverberation mapping sample
  range. It is also possible that the scatter about the $M_{BH}$--$L$
  relationship depends on $z$.  The most likely source of a redshift
  dependence for $\epsilon_l$ is evolution of the distribution of
  $\dot{m}$ \citep[e.g.,][]{hai00,merloni04,hop06a,steed06}, and
  possibly evolution of the bolometric correction. The average
  $\dot{m}$ has been observed to increase with increasing $z$
  \citep{mclure04a}, but this is likely due to selection
  effects. Because of these issues, one should exhibit caution when
  applying this $M_{BH}$--$L$ relationship to high $z$ sources.

  \subsection{Inferring the C IV $R$--$L$ Relationship}

  \label{s-civrl}

  In this subsection, we investigate the C IV $R$--$L$ relationship
  using the reverberation mapping sample. We perform a linear
  regression to fit for the value of $\theta_l, R \propto
  L^{\theta_l}$, by noting that $M_{BH} \propto L^{\theta_l}
  V^2$. Based on the results obtained here, we find $M_{BH} \propto
  L_{1549}^{0.41 \pm 0.07} FWHM_{CIV}^2$, consistent with $M_{BH} \propto
  L_{1549}^{1/2} FWHM_{CIV}^2$ expected from simple photoionization
  theory.

  To estimate $\theta_l$, we fit a linear relationship of the form $2v
  = m - m^{BL}_0 - \theta_l l$ using the BCES($Y|X$) method. Here, the
  free parameters are $m^{BL}_0$ and $\theta_l$. The result is
  \begin{equation}
    \hat{m}_{CIV} = -17.82 (\pm 2.99) + 0.41 (\pm 0.07) l + 2 \log FWHM_{CIV}.
    \label{rlfit}
  \end{equation}
  The intrinsic scatter about this relationship is $\hat{\sigma}_{CIV}
  = 0.40$ dex, where we have corrected for the measurement errors in
  $m$ and $FWHM_{CIV}$. The correlation between the regression
  coefficients is $Corr(\hat{m}^{CIV}_0, \hat{\theta}_l) =
  -0.9996$. The regression results are consistent with the expectation
  from simple photoionization theory, $R \propto L^{1/2}_{1549}$
  (cf. Eq.[\ref{ionpar}]). A Kolmogorov-Smirnov test found that the
  regression residuals are not significantly different from a normal
  distribution.

  Performing the regression with $v$ as the dependent variable ensures
  that the $M_{BH}$--$L$ relationship does not `absorb' into the
  regression coefficients, as the regression models the distribution
  of $v$ at a given $l$ and $m$. However, if we had performed the
  regression by fitting the distribution of $m$ as a function of $l$
  and $v$, the intrinsic correlation between $m$ and $l$ would have
  absorbed into the results. In this case we would be modelling
  $p(m|l,v)$, and the statistical model would be over-parameterized
  because we have only three variables. However, by seperating
  $p(m|l,v)$ into $p(l|m)$ and $p(v|l,m)$, as was done in
  \S~\ref{s-sespec}, we can analyze each distribution seperately and
  uniquely determine their parameters.

  The residuals of the C IV broad line estimate, given by Equation
  (\ref{rlfit}), and their CDF are shown in Figure
  \ref{f-photoresid}. In general, there is no obvious evidence for a
  violation of the regression assumptions. However, there appears to
  be a possible trend in the residuals with $FWHM$. To test this we
  calculated the linear correlation coefficient between the residuals
  and $\log FWHM$. This correlation was significant at only $\approx
  1.7\sigma$. Therefore there is no significant evidence for a
  correlation in the residuals with $FWHM$, justifying the assumption
  $M_{BH} \propto L_{1549}^{\theta_l} FWHM^2$. Rank correlation tests
  gave similar results.

  Very similar results were found by \citet{vest06}. However, our work
  differs from their's in that \citet{vest06} assumed that the C IV and
  H$\beta$ BLRs have the same dependence on $L_{UV}, R \propto
  L_{UV}^{0.53}$, and only fit the constant term. In contrast, we fit
  both the constant term and the coefficient for the dependence of
  line width on luminosity.

  One can use Equation (\ref{wmean}) to combine the mass estimates
  based on the $M_{BH}$-$L$ with the C IV broad line mass
  estimate. Combining the two relationships, and taking $\sigma(z)
  \rightarrow \infty$, we find
  \begin{equation}
    \mu = -30.53 + 0.54 l + 1.40 \log FWHM.
    \label{mvest}
  \end{equation}
  The intrinsic uncertainty in $\mu$ is reduced to $\sigma_{\mu}
  \approx 0.33$ dex, an improvement of $\approx 18\%$ over the broad
  line mass estimates. However, as mentioned in \S~\ref{s-mlrel},
  there is considerable systematic uncertainty on the behavior of the
  $M_{BH}$--$L$ relationship outside of the range probed by the
  reverberation mapping sample, and thus it may be safer to just use
  $\hat{m}_{CIV}$.

  There has been some concern over the effectiveness of using the C IV
  emission line for estimating quasar black hole masses \citep{shem04,
  bask05}. If one ignores the C IV $FWHM$, and only uses the continuum
  luminosity to estimate $M_{BH}$, then the intrinsic scatter is
  $\sigma_{l} \approx 0.6$ dex (cf. \S~\ref{s-mlrel}). However, the
  intrinsic uncertainty in an estimate of $M_{BH}$ using the C IV
  $R$--$L$ relationship and $FWHM$ is $\sigma_{CIV} \approx 0.4$ dex,
  a improvement over the $M_{BH}$--$L$ relationship of $\sim
  1/3$. This improvement is significant, and therefore we conclude
  that the C IV line may be used for estimating SMBH masses.

  We can estimate $\sigma_v^2$ by comparing the C IV $FWHM$ for the
  RMS spectrum with that obtained from a single-epoch spectrum. To do
  this we use the sources from \citet{peter04} with C IV $FWHM$
  measured from the RMS spectrum, $FWHM_{RMS}$, excluding Fairall
  9. We omitted Fairall 9 because its $FWHM_{RMS}$ was considered to
  be unreliable by \citet{peter04}. The remaining five data points are
  consistent with a 1:1 relationship between $FWHM_{RMS}$ and
  $FWHM_{SE}$. Estimating the intrinsic scatter using these five data
  points is difficult, particularly because the measurement errors are
  large. In fact, the observed scatter about $FWHM_{RMS} = FWHM_{SE}$
  is consistent with entirely being the result of the measurement
  errors. This suggests that $\sigma_v$ is small, and it is likely
  that $\sigma_v \lesssim 0.1$ dex.

  If we assume that the only sources for scatter in the broad line
  estimates of $m$ are from the $R$--$L$ relationship and from using
  the single-epoch line width to estimate $V$, then a value of
  $\sigma_v \approx 0.1$ dex implies that the intrinsic scatter about
  the $R$--$L$ relationship for C IV is $\hat{\sigma}_r \approx 0.35$
  dex. This is about a factor of two larger than the value
  $\hat{\sigma}_r \approx 0.18$ dex found for H$\beta$ by
  \citet{peter04}, where we have converted the value $\hat{\sigma}_r
  \approx 40\%$ to dex. This larger scatter in the C IV $R$--$L$
  relationship implies that the H$\beta$ mass estimates are more
  efficient than the C IV ones. However, \citet{vest06} found that the
  H$\beta$ mass estimates have a statistical scatter of
  $\hat{\sigma}_{{\rm H}\beta} \approx 0.43$ dex, and thus are just as
  accurate as the C IV-based ones. This larger value of $\sigma_{{\rm
  H}\beta}$ implies that either there is significant uncertainty in
  using the SES H$\beta$ line width to estimate the BLR velocity
  dispersion, $\sigma_v^{{\rm H}\beta} \approx 0.36$, or that there
  are additional sources of uncertainty in the broad line mass
  estimates beyond the scatter in the $R$--$L$ relationship and the
  uncertainty in using the SES line width instead of the variable
  component line width. If there are additional sources of scatter in
  the SES mass estimates, then $\hat{\sigma}_r \approx 0.35$ dex
  represents an upper bound on the C IV $R$--$L$ relationship scatter.

  Equation (\ref{ionpar}) implies that the scatter in the $R$--$L$
  relationship is the result of variations in the BLR ionization
  parameter, gas density, and average ionizing photon energy. While
  there are likely other sources of scatter in the $R$--$L$
  relationship, such as inclination, variability, and conversion
  between $L_{\lambda}$ and $L_{ion}$, differences in BLR properties
  certainly contribute to this scatter and possibly dominate it. The
  fact that $\sigma^{CIV}_r \lesssim 1.6 \sigma^{{\rm H}\beta}_r$
  suggests that the magnitude of the dispersion in these properties is
  larger for the C IV emitting region. In addition, the regressions of
  \citep{peter04} are consistent with $R_{{\rm H}\beta} \propto
  L_{1450}^{1/2}$, similar to the result found here and by
  \citet{vest06} for C IV.

  \section{COMPARING THE H$\beta$- AND C IV-BASED MASS ESTIMATES}
  
  \label{s-hbeta}

  It is worth comparing our estimates of $m$ based on single-epoch C
  IV to those based on combining single-epoch H$\beta$ with the
  $R_{{\rm H}\beta}$--$L$ relationship. To do this, we took our sample
  of 100 sources that have data for both C IV and H$\beta$ and
  calculated estimates of $m$ from each. We compare the mass estimates
  obtained from both emission lines, and find that the two give mass
  estimates that are consistent so long as one uses the UV continuum
  to estimate the BLR size.
  
  For the H$\beta$ emission line, we estimate the BLR size from both
  the optical and UV $R_{{\rm H}\beta}$--$L$ relationships. The BLR
  size, $R_{{\rm H}\beta}$, is estimated as
  \begin{equation}
    \frac{R_{{\rm H}\beta}}{10 \ {\rm lt-dy}} = R_0 \left( \frac{\lambda
      L_{\lambda}}{10^{44} {\rm ergs \ s}^{-1}}
    \right)^{\alpha}.
    \label{radlum}
  \end{equation}
  For $R_0$ and $\alpha$ in the optical $R_{{\rm H}\beta}$--$L$
  relationship, we use the luminosity at $5100\AA$ and the average of
  the BCES bisector and FITEXY \citep{numrec} fits of \citet{kaspi05},
  $(\hat{R}_0, \hat{\alpha}) = (2.23 \pm 0.21, 0.69 \pm 0.05)$.  For
  the UV $R_{{\rm H}\beta}$--$L$ relationship, we use the luminosity
  at $1450\AA$ and $(R_0, \alpha) = (2.38, 0.5)$. We assume $R_{{\rm
  H}\beta} \propto L_{1450}^{1/2}$ to allow more direct comparison
  with the C IV-based mass estimates, and because the fits of
  \citet{kaspi05} are consistent with this form. Using the averaged
  values of $R$ for the Balmer lines as listed by \citet{kaspi05}, we
  recalculated the $R$--$L_{1450}$ relationship for H$\beta$ with
  $\alpha$ fixed at $\alpha = 0.5$, and found $R_0 = 2.38 \pm
  0.25$. We then use $R_{{\rm H}\beta}$ with H$\beta$ $FWHM$ to
  calculate $m_{{\rm H}\beta}$ from Equation (\ref{virial}), after
  converting the H$\beta$ $FWHM$ to a velocity dispersion assuming
  $FWHM / V = 2$, based on the average $FWHM / \sigma_*$ from
  \citet{peter04}.

  Based on the results of \S~\ref{s-civrl}, we calculate estimates of
  $M_{BH}$ from the C IV line for these sources as $\hat{m}_{CIV} =
  -21.92 + 0.5 l_{1549} + 2 v_{CIV}$. This form was found using the
  same method as described in \S~\ref{s-civrl}, but with $\theta_l$
  fixed at $\theta_l = 1/2$. In Table \ref{t-mbhest} we report
  the broad line mass estimates based on the H$\beta$ and C IV lines
  for these sources, as well as a weighted average of the two.

  In Figure \ref{f-mhb_vs_mciv} we show $\hat{m}_{{\rm H}\beta}$ vs
  $\hat{m}_{\rm CIV}$ for both the optical and UV $R_{{\rm
  H}\beta}$--$L$ relationship. We also show the $99\%$ ($2.6\sigma$)
  confidence region for the parameters of a BCES bisector fit to
  $\hat{m}_{CIV} = A + B \hat{m}_{{\rm H}\beta}$. As can be seen, the
  values of $\hat{m}_{{\rm H}\beta}$ calculated using the UV $R_{{\rm
  H}\beta}$--$L$ relationship are more consistent with the C IV-based
  ones. In addition, a pure 1:1 relationship falls outside of the
  $99\%$ confidence region on the BCES bisector parameters for the
  optical-based $\hat{m}_{{\rm H}\beta}$ values, but it is contained
  within the $99\%$ confidence region for the UV-based estimates. In
  fact, the BCES bisector fit for the UV-based mass estimates differs
  from a 1:1 relationship at a significance level of only
  $1.9\sigma$. Our results are in agreement with \citet{warn03} and
  \citet{diet04}, who also compared the C IV- and H$\beta$-based mass
  estimates and found them to be consistent.

  That the C IV- and H$\beta$-based mass estimates are consistent is
  in contrast to the conclusions of \citet{bask05} and
  \citet{shem04}. \citet{bask05} suggested that C IV-based mass
  estimates may be problematic because they find that $\log (FWHM_{\rm
  C IV} / FWHM_{{\rm H}\beta})$ is significantly anti-correlated with
  $\log FWHM_{{\rm H}\beta}$. However this is expected even when
  $v_{{\rm H}\beta} \propto v_{\rm CIV}$, as an anti-correlation
  between the ratio of the line widths and the H$\beta$ $FWHM$ is
  expected by construction; i.e., an anti-correlation between $\log
  FWHM_{\rm C IV} - \log FWHM_{{\rm H}\beta}$ and $\log FWHM_{{\rm
  H}\beta}$ is expected so long as the deviations of $\log FWHM_{\rm C
  IV}$ and $\log FWHM_{{\rm H}\beta}$ from there respective means are
  not strongly correlated. In addition, none of the sources in the
  \citet{bask05} sample have H$\beta$ broader than C IV for
  $FWHM_{{\rm H}\beta} < 4000 {\rm\ km\ s^{-1}}$. To a lesser extant,
  a similar effect is seen in Figure 7 of \citet{shem04} and Figure 4
  of \citet{warn03}. In Figure \ref{f-mhb_vs_mciv} we compare the C IV
  and H$\beta$ $FWHM$ for our sources. We find that most of the
  sources with $FWHM_{{\rm H}\beta} \lesssim 2000 {\rm \ km\ s^{-1}}$
  tend to have broader C IV lines. While the line widths certainly do
  not have a 1:1 relationship, the BCES bisector fit found that on
  average $FWHM_{{\rm H}\beta}$ is approximately proportional to
  $FWHM_{CIV}$, $FWHM_{CIV} \propto FWHM_{{\rm H}\beta}^{0.79 \pm
  0.06}$. \citet{bask05} and \citet{shem04} did not perform a
  regression analysis, so we are unable to make a more quantitative
  comparison.
 
  If the line widths are set by the virial relationship, we would
  expect $FWHM_{CIV} \propto FWHM_{{\rm H}\beta}$. While the
  divergence from $FWHM_{CIV} \propto FWHM_{{\rm H}\beta}$ is likely
  real ($3.5\sigma$ significance), it is small, especially when
  compared with the intrinsic scatter in the $FWHM$ plot. Furthermore,
  this divergence from $FWHM_{CIV} \propto FWHM_{{\rm H}\beta}$ does
  not appear to have a significant effect on the C IV mass estimates,
  as the masses inferred from the two emission lines differ only at
  the level of $1.9\sigma$. As can by seen from the plot in Figure
  \ref{f-mhb_vs_mciv}, any systematic difference between the C IV- and
  H$\beta$-based mass estimates is small compared to the statistical
  scatter in $\hat{m}$.

  \citet{vest06} performed a reanalysis of the \citet{bask05} sample
  and concluded that the poor correlation between the C IV $FWHM$ and
  the H$\beta$ $FWHM$ seen by \citet{bask05} was due to the inclusion
  of a large number of Narrow Line Seyfert 1s (NLS1) and the
  subtraction of a C IV narrow component. \citet{vest04} noted that
  for these sources the C IV line profile is unlikely to be suitable
  for estimating $M_{BH}$, as there may be a strong component from an
  outflowing wind. Upon removing the NLS1s and the sources with poor
  IUE data, \citet{vest06} found the correlation between C IV $FWHM$
  and H$\beta$ $FWHM$ to be significantly better.

  In addition, \citet{bask05} compared the correlation of C IV $EW$
  with $L / L_{edd} \propto L / M_{BH}$. A similar investigation was
  performed by \citet{shem04}, where they compared the metallicity
  indicator N V/C IV with $L / L_{edd}$. Both authors found a stronger
  correlation when $M_{BH}$ was estimated from the H$\beta$ line as
  compared to C IV, and suggested that the C IV line may give a less
  efficient and possibly biased estimate of $M_{BH}$.  However, as
  noted in \S~\ref{s-corrprob}, correlation coefficients inferred from
  the estimated black hole masses must be interpreted with caution. In
  particular, the SES estimate of $m$, $\hat{m}$, merely defines the
  centroid of the probability density of $m|l,v,z$, and is in general
  {\em not} the actual $m$ for an object. Therefore, this difference
  in the correlation coefficient between the H$\beta$- and C IV-based
  mass estimates may be just the result of random sampling, and it is
  unclear whether one can say that the correlation coefficients
  inferred from the two different mass estimates are inconsistent with
  those mass estimates being drawn from the same parent distribution.

  The most direct test of the effectiveness of using C IV to estimate
  $M_{BH}$ is found by comparing those values of $\hat{m}$ estimated
  using C IV with the actual reverberation mapping values of $m$. This
  was done in \S~\ref{s-civrl}, where inclusion of the C IV $FWHM$
  resulted in a reduction of the error in $\hat{m}$ of $\sim 1/3$. In
  addition, Figure \ref{f-photoresid} compares the C IV-based estimate
  of $m$ with the reverberation mapping values. As can be seen,
  assuming $M_{BH} \propto FWHM^2_{CIV}$ is consistent with the
  reverberation mapping sample, and therefore there is no reason to
  assume that the C IV line gives a biased estimate of $M_{BH}$. If
  $\hat{m}_{CIV}$ does give a biased estimate of $m$, then this bias
  is likely negligible compared to the variance in $\hat{m}_{CIV}$.

  \section{SUMMARY}

  \label{s-summary}

  In this work we have undertaken a statistical investigation of the
  best method to estimate SMBH mass based on the single-epoch C IV
  line and AGN continuum at $1549\AA$. The main conclusions are :
  \begin{enumerate}
  \item
    Estimating AGN black hole masses from a single-epoch spectrum is a
    considerably different problem than in the reverberation-mapping
    case. Because of this, one is not estimating $M_{BH}$ directly,
    but calculating a probability distribution of $M_{BH}$ given the
    observed luminosity, line width, and redshift, $p(m|l,v,z) \propto
    p(v|l,m,z) p(l|m,z) p(m|z)$.
  \item
    Combining the information in $L$ from both an intrinsic
    $M_{BH}$--$L$ correlation and the $R$--$L$ relationship results in
    improved black hole mass estimates. However, because of the
    current systematic uncertainties in the $M_{BH}$--$L$
    relationship, estimates based on its inclusion should be viewed
    with caution. In addition, incorporating information from the
    intrinsic distribution of $M_{BH}$ also results in more accurate
    estimates, on average.
  \item
    The distribution of $M_{BH}$ inferred from the broad line mass
    estimates, $\hat{m}_{BL}$, is broader than the intrinsic
    distribution of $M_{BH}$, as $\hat{m}_{BL}$ are estimates of $m$
    contaminated by `measurement error.' In addition, it is necessary
    to estimate or assume $p(l|m,z)$ in order to estimate a survey's
    completeness in $M_{BH}$.
  \item 
    The best estimates of $M_{BH}$ based on the C IV emission line are
    obtained using the UV continuum luminosity and $FWHM$. We find
    evidence that the C IV line shift, centroid, $EW$, and spectral
    slope of the UV continuum do not contribute any additional
    information to estimating $M_{BH}$ for a given $L_{1549}$ and
    $FWHM_{CIV}$.
  \item
    Using the reverberation mapping sample, we find an $M_{BH}$--$L$
    relationship of the form
    \begin{equation}
      l = 35.72(\pm 1.67) + 1.17(\pm 0.22) m \nonumber.
    \end{equation}
    The intrinsic scatter about this relationship is $\sigma_l \approx
    0.61$ dex. Combining mass estimates based on this relationship
    with the broad line mass estimates results in a reduction in the
    statistical error of $\approx 18\%$.
  \item
    We estimate a C IV $R$--$L$ relationship of the form $R \propto
    L_{1549}^{0.41 \pm 0.07}$. This is consistent with an $R$--$L$
    relationship of the form $R \propto L^{1/2}$ predicted from simple
    photoionization physics or if the BLR size is set by the dust
    sublimation radius. The scatter about the C IV $R$--$L$
    relationship inferred from the reverberation mapping estimates of
    $M_{BH}$ is $\sigma_r \approx 0.35$ dex.
  \item
    A broad line estimate of $m$ based on the C IV emission line may
    be calculated as
    \begin{equation}
      \log \hat{M}_{BH} / M_{\odot} = \hat{m}_{CIV} = -17.82 (\pm 2.99) +
      0.41 (\pm 0.07) \log \lambda L_{\lambda}(1549\AA) + 2 \log FWHM_{CIV}.
      \label{mciv}
    \end{equation}
    Here, $\lambda L_{\lambda}$ is in units of ${\rm ergs\ s^{-1}}$
    and $FWHM$ is in units of ${\rm km\ s^{-1}}$. The correlation
    between regression coefficients is $Corr(\hat{m}^{CIV}_0,
    \hat{\theta}_l) = -0.9996$. The intrinsic scatter in $m$ about
    Equation (\ref{mciv}) is $\approx 0.40$ dex.
  \item
    The C IV- and H$\beta$-based mass estimates are consistent if one
    assumes $R \propto L_{UV}^{1/2}$ for both emission lines. The two
    emission lines give estimates of $M_{BH}$ with comparable
    accuracy.
  \item
    We find evidence that the $R$--$L$ relationships for C IV and
    H$\beta$ are similar in their dependence on $L_{UV}$. For both, $R
    \propto L_{UV}^{1/2}$ is consistent with the available data. Also,
    the scatter in the C IV $R$--$L$ relationship is a factor
    $\lesssim 1.6$ larger. This suggests that the C IV broad line
    region gas has a larger dispersion in its properties (e.g.,
    density, ionization parameter, inclination) than the H$\beta$ BLR
    gas.
  \end{enumerate}

  \acknowledgements
  
  We are grateful to Marianne Vestergaard for providing the UV Fe
  emission template used in this study, Jun Cui for directly obtaining
  the spectra for Q 1230+0947, and Marianne Vestergaard, Xiaohui Fan,
  Fulvio Melia, and Aneta Siemiginowska for stimulating discussions
  and helpful comments on an earlier version of this manuscript. We
  are also grateful to the referee for a careful reading of this work
  and helpful comments that contributed to the significant improvement
  of this paper. This research was supported in part by NASA/Chandra
  grant G04-5112X. This paper includes data gathered with the 6.5
  meter Magellan Telescopes located at Las Campanas Observatory,
  Chile.
  
  This research has made use of the NASA/IPAC Extragalactic Database
  (NED) which is operated by the Jet Propulsion Laboratory, California
  Institute of Technology, under contract with the National
  Aeronautics and Space Administration.

  Funding for the SDSS and SDSS-II has been provided by the Alfred
  P. Sloan Foundation, the Participating Institutions, the
  National Science Foundation, the U.S. Department of Energy, the
  National Aeronautics and Space Administration, the Japanese
  Monbukagakusho, the Max Planck Society, and the Higher Education
  Funding Council for England. The SDSS Web Site is
  http://www.sdss.org/.
  
  The SDSS is managed by the Astrophysical Research Consortium for
  the Participating Institutions. The Participating Institutions are
  the American Museum of Natural History, Astrophysical Institute
  Potsdam, University of Basel, Cambridge University, Case Western
  Reserve University, University of Chicago, Drexel University,
  Fermilab, the Institute for Advanced Study, the Japan
  Participation Group, Johns Hopkins University, the Joint Institute
  for Nuclear Astrophysics, the Kavli Institute for Particle
  Astrophysics and Cosmology, the Korean Scientist Group, the
  Chinese Academy of Sciences (LAMOST), Los Alamos National
  Laboratory, the Max-Planck-Institute for Astronomy (MPIA), the
  Max-Planck-Institute for Astrophysics (MPA), New Mexico State
  University, Ohio State University, University of Pittsburgh,
  University of Portsmouth, Princeton University, the United States
  Naval Observatory, and the University of Washington.

  \clearpage
  
  \setcounter{figure}{2}

  \begin{figure}
    \begin{center}
      \figurenum{2a}
      \scalebox{1.0}{\rotatebox{0}{\plotone{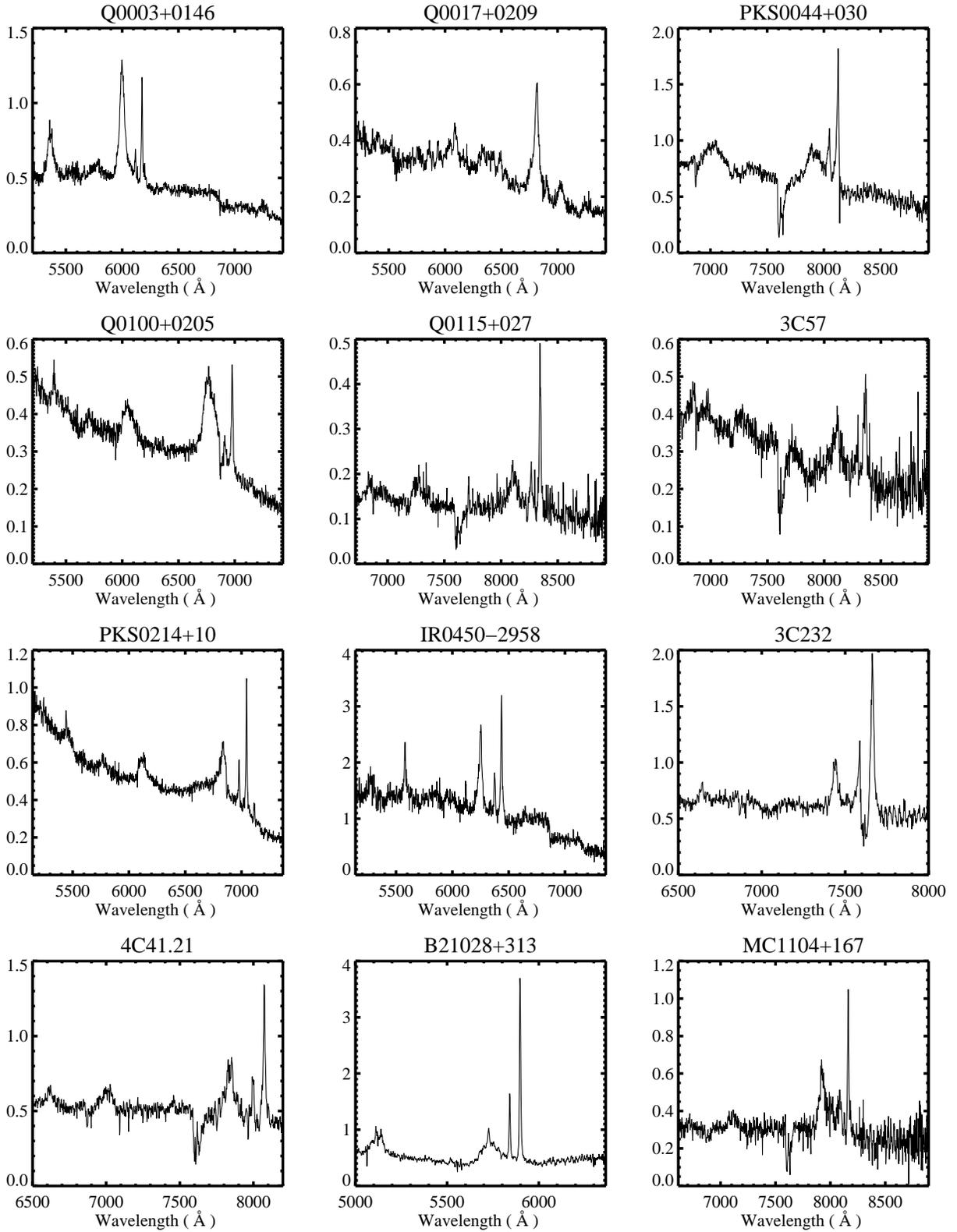}}}
      \caption{New optical spectra, shown in the observed frame. The
      fluxes are in units of $10^{-15}$ ergs cm$^{-2}$ sec$^{-1}$
      $\AA^{-1}$. The absorption features at $\sim 6875 \AA$ and $\sim
      7600 \AA$ are the A- and B-band atmospheric absorption
      lines. Note the strong iron emission in
      Q1340-0038. \label{f-newspec}}
    \end{center}
  \end{figure}

  \clearpage

  \begin{figure}
    \begin{center}
      \figurenum{2b}
      \scalebox{1.0}{\rotatebox{0}{\plotone{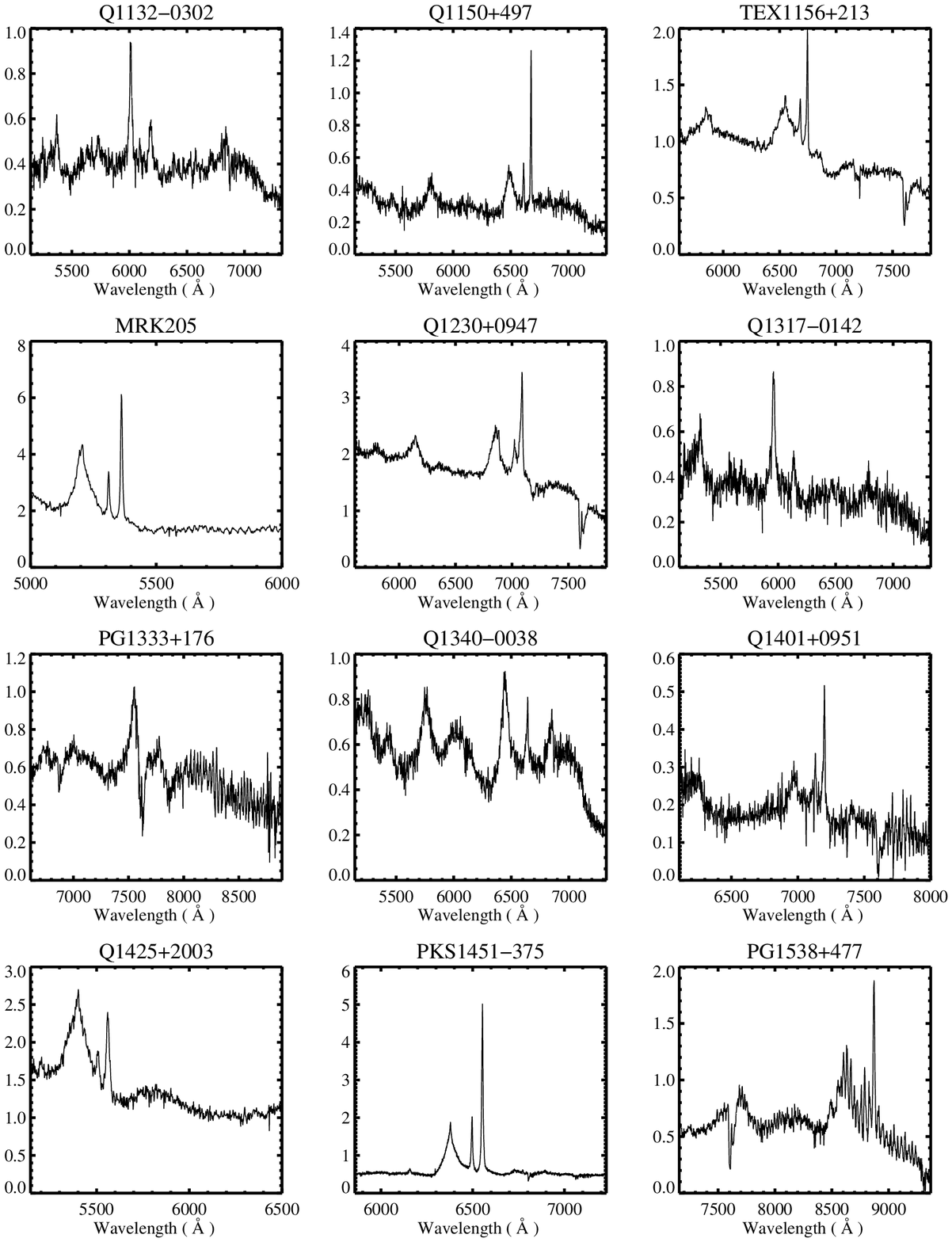}}}
      \caption{}
    \end{center}
  \end{figure}

  \clearpage

  \begin{figure}
    \begin{center}
      \figurenum{2c}
      \scalebox{1.0}{\rotatebox{0}{\plotone{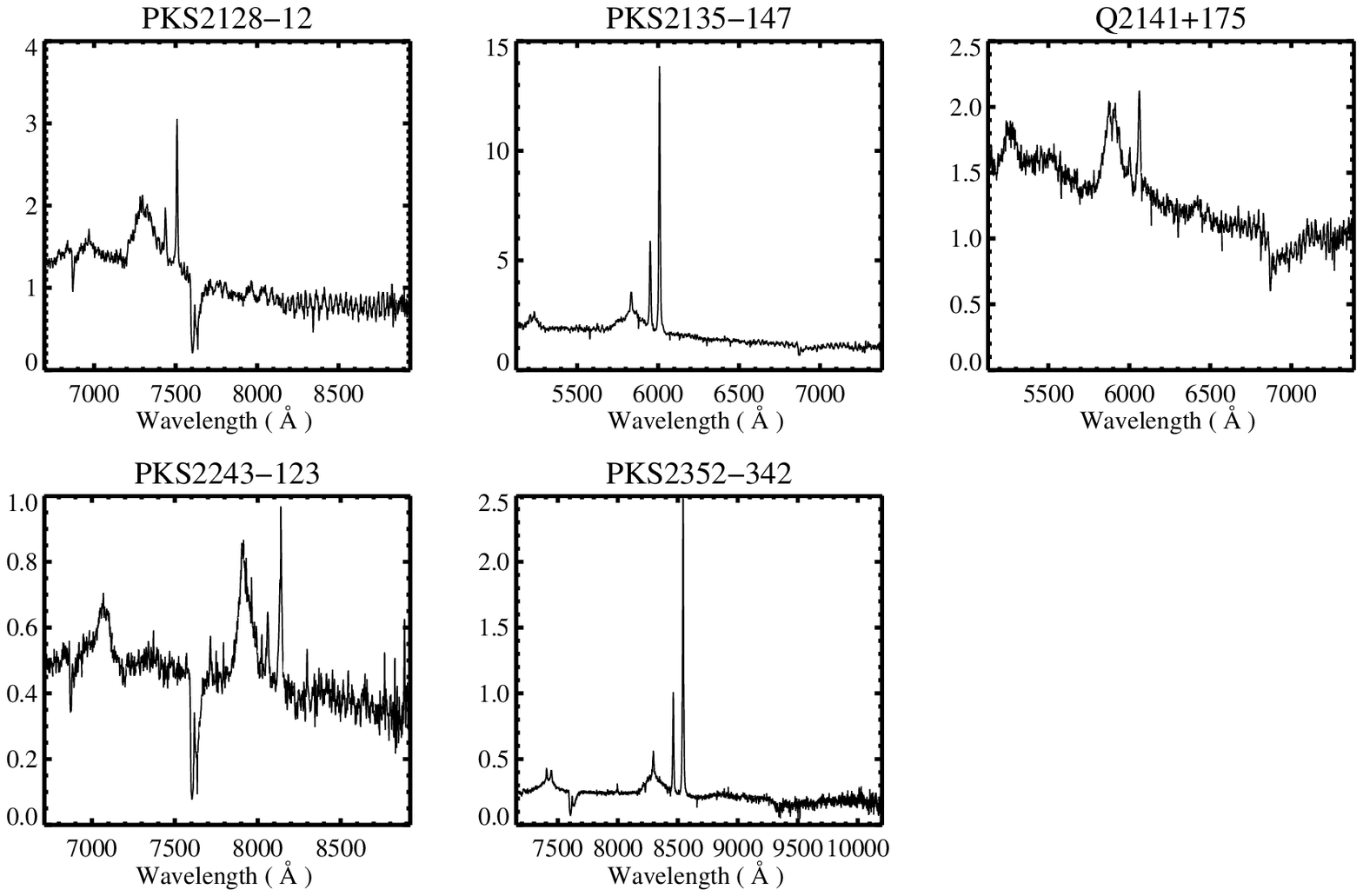}}}
      \caption{}
    \end{center}
  \end{figure}

  \clearpage

  \setcounter{figure}{2}

  \begin{figure}
    \begin{center}
      \scalebox{0.8}{\rotatebox{90}{\plotone{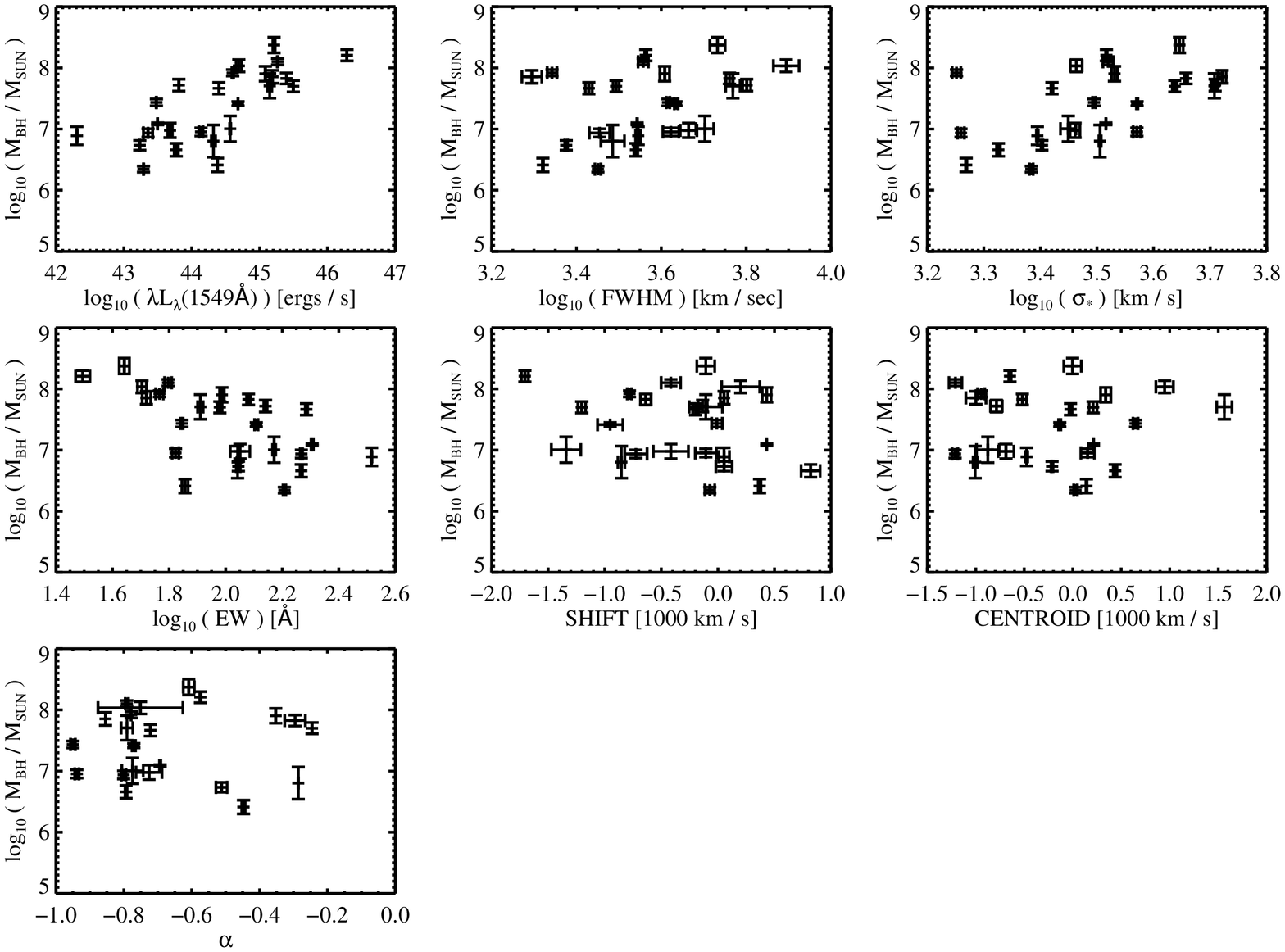}}}
      \caption{Plots showing $m$ as a function of the measured
      quantities for the sample with reverberation mapping data. Three
      outliers in the $m$ vs $\alpha$ plot have been removed so as to
      make the structure in the plot easier to see. \label{f-mvx}}
        \end{center}
  \end{figure}

  \clearpage
  
  \begin{figure}
    \begin{center}
      \includegraphics[scale=0.4,angle=90]{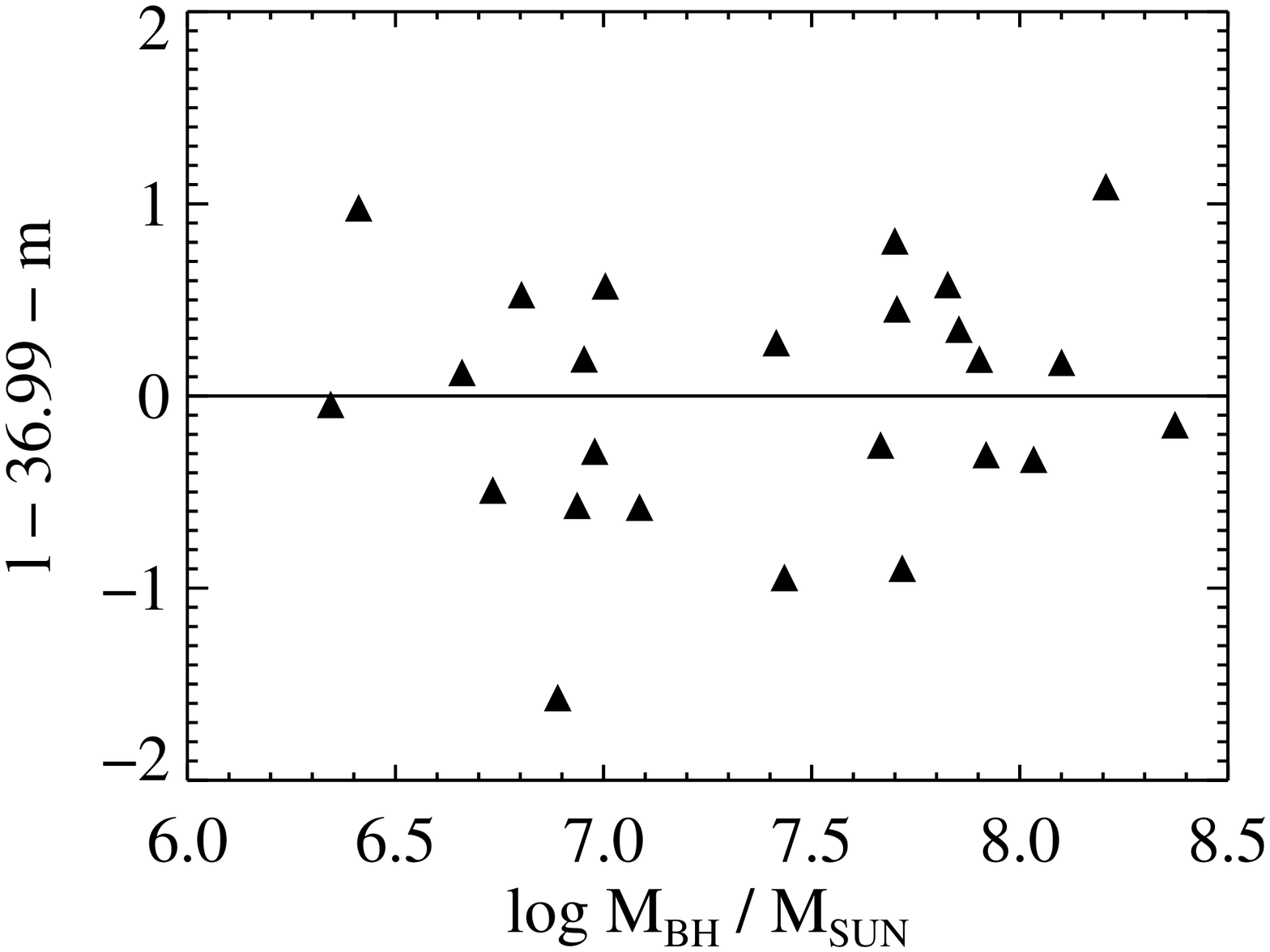}
      \includegraphics[scale=0.4,angle=90]{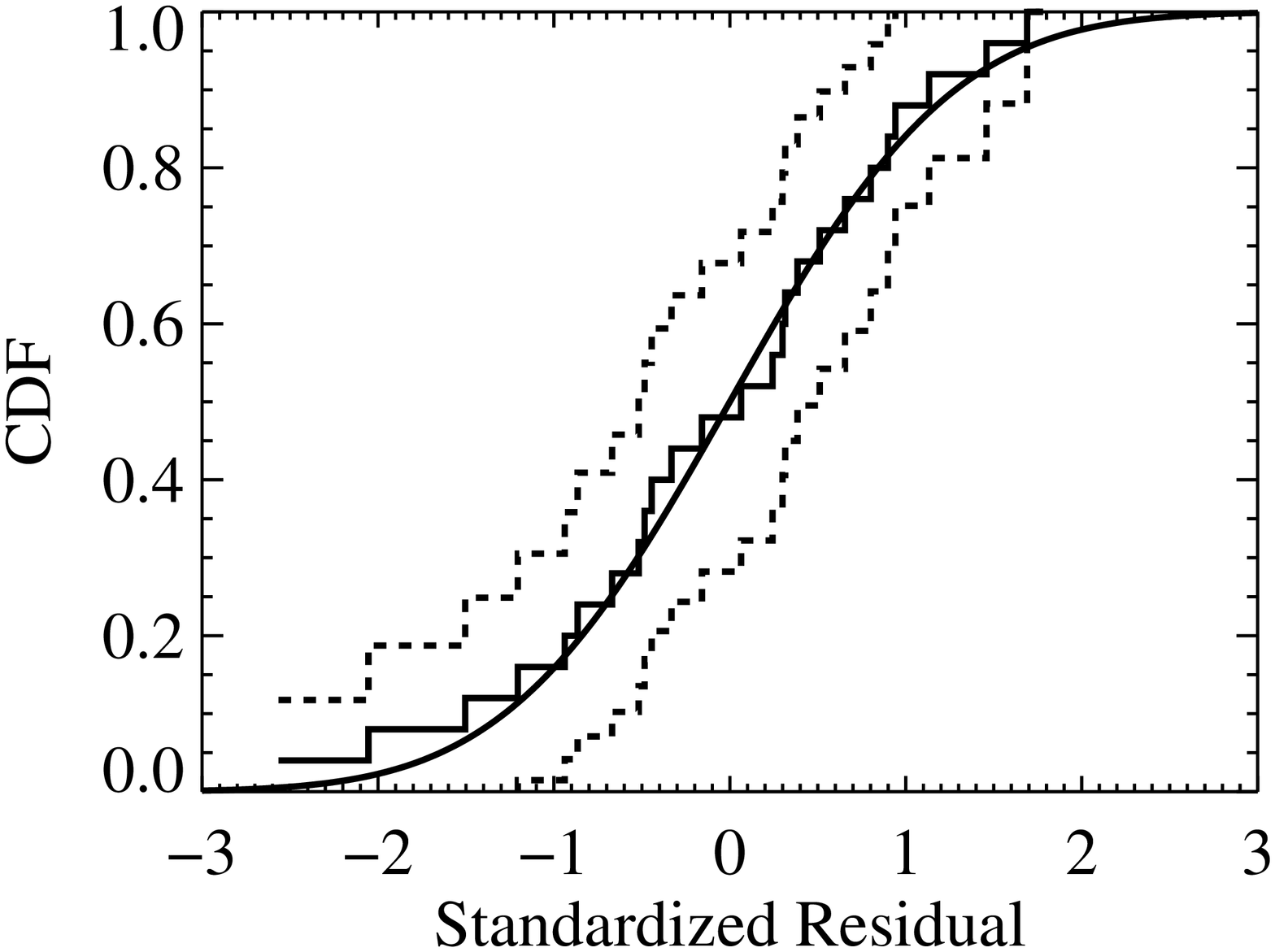}
      \caption{The residuals $l - \hat{l}$ as a function of $m$ for
      the $M_{BH}$--$L$ relationship and the empirical CDF of the
      standardized residuals, $(l - \hat{l}) / \hat{\sigma}_l$. The
      dashed lines define the $95\%$ pointwise confidence interval of
      the empirical CDF, and the smooth solid line is the CDF of the
      standard normal density.\label{f-mlumresid}}
    \end{center}
  \end{figure}

  \clearpage

  \begin{figure}
    \begin{center}
      \scalebox{0.75}{\rotatebox{90}{\plotone{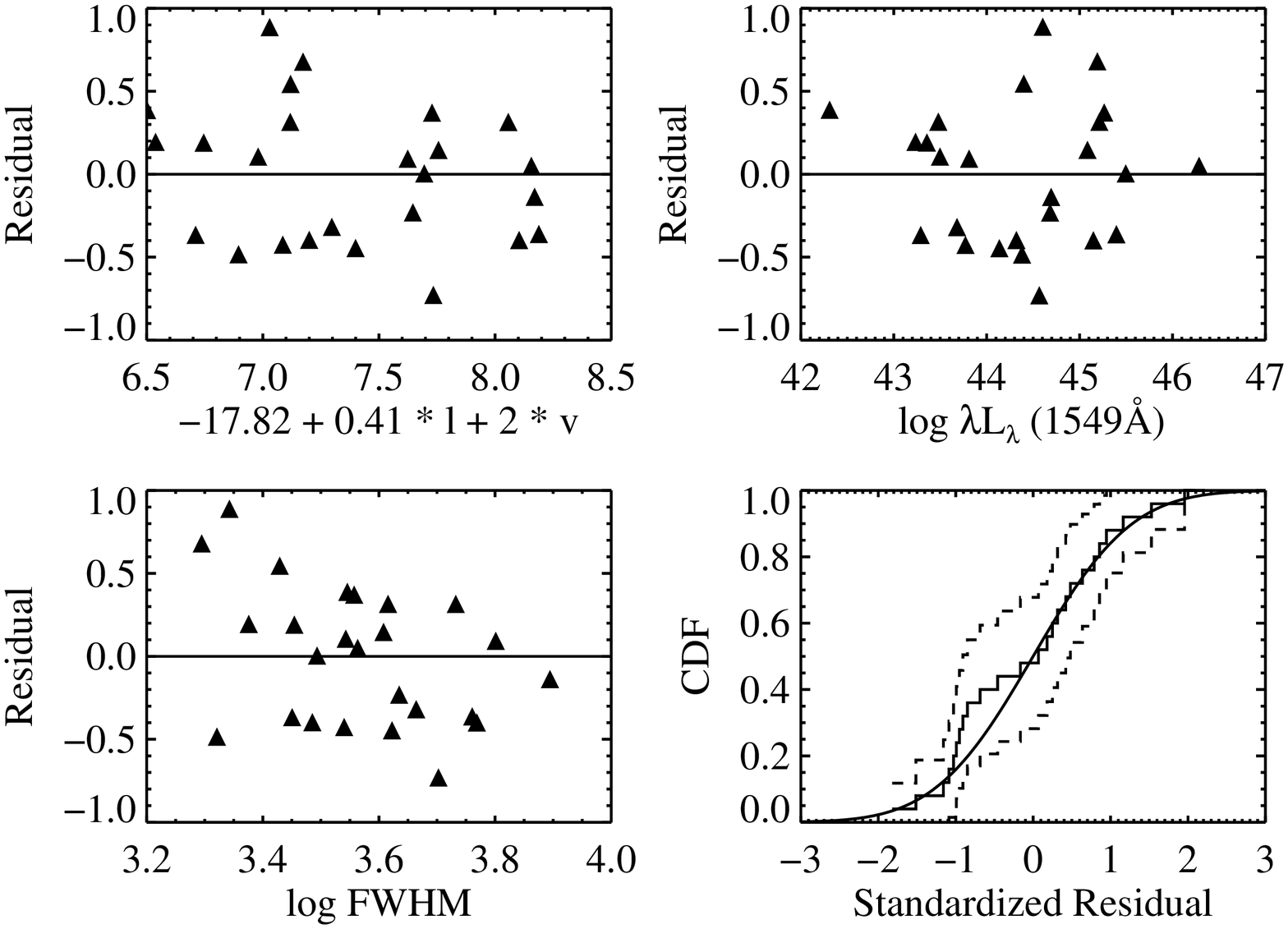}}}
      \caption{The residuals $m - \hat{m}_{CIV}$ for the C IV broad
      line mass estimate, shown as a function of $\hat{m}_{CIV}, l$,
      and $FWHM$. Also shown is the empirical CDF of the standardized
      residuals, $(m - \hat{m}_{CIV}) / \hat{\sigma}_{CIV}$. The
      dashed lines define the $95\%$ pointwise confidence interval of
      the empirical CDF, and the smooth solid line is the CDF of the
      standard normal density.\label{f-photoresid}}
    \end{center}
  \end{figure}
 
  \clearpage
  
  \begin{figure}
    \begin{center}
      \scalebox{0.75}{\rotatebox{90}{\plotone{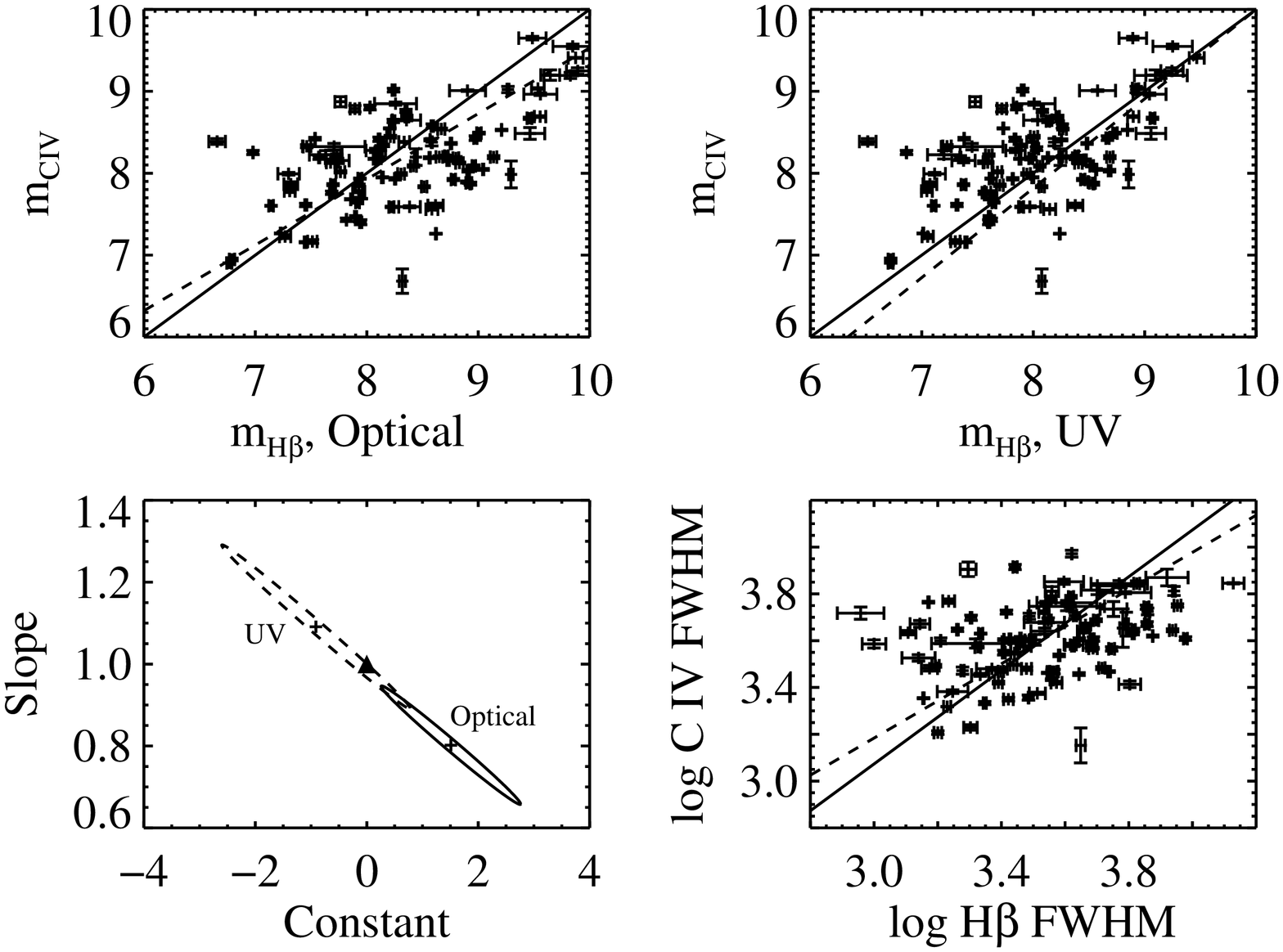}}}
      \caption{Plots comparing the H$\beta$-based mass estimates with
      the C IV-based ones. The values of $\hat{m}_{{\rm H}\beta}$ for
      the top left plot are calculated from the optical $R$--$L$
      relationship, while the values of $\hat{m}_{{\rm H}\beta}$ for
      the top right plot assume $R_{{\rm H}\beta} \propto
      L_{1450}^{1/2}$. The error bars denote the propagated
      measurement errors in $\hat{m}$ resulting from the measurement
      errors in the line widths and continuum luminosities. The solid
      lines are a 1:1 relationship and the dashed lines are the BCES
      bisector fits. The bottom left plot shows the $99\%$
      ($2.6\sigma$) confidence regions for the BCES bisector fits
      comparing $\hat{m}_{CIV}$ with $\hat{m}_{{\rm H}\beta}$. The
      solid contour correspond to $\hat{m}_{{\rm H}\beta}$ calculated
      from the optical $R_{{\rm H}\beta}$--$L$ relationship, and the
      dashed contour when $\hat{m}_{{\rm H}\beta}$ is calculated from
      the UV one. The crosses mark the best-fit values, and the
      triangle marks the values expected for a 1:1 relationship. The
      bottom right plot shows a comparison between the H$\beta$ and C
      IV $FWHM$. The solid line is the best fit $FWHM_{{\rm H}\beta}
      \propto FWHM_{CIV}$ relationship, and the dashed line is the
      BCES bisector fit.\label{f-mhb_vs_mciv}}
    \end{center}
  \end{figure}

  \clearpage

  \begin{deluxetable}{lcccccc}
    \tabletypesize{\scriptsize}
    \tablecaption{Object List\label{t-data}}
    \tablewidth{0pt}
    \tablehead{
      \colhead{Object} & \colhead{RA (J2000)} & \colhead{DEC (J2000)} & \colhead{Redshift} & 
      \colhead{$\log (M_{BH} / M_{\odot})$\tablenotemark{a}} & \colhead{Rad. Type\tablenotemark{b}} & 
      \colhead{Instrument}
    }
    \startdata
    MRK 335     & 00 06 19.5 & +20 12 10.5 & 0.025  & 6.41 $\pm$ 0.11 & Quiet & FOS \\
    PG 0026+129 & 00 29 13.7 & +13 16 03.8 & 0.142  & 7.85 $\pm$ 0.10 & Quiet & FOS \\
    PG 0052+251 & 00 54 52.1 & +25 25 39.3 & 0.155  & 7.82 $\pm$ 0.08 & Quiet & FOS \\
    FAIRALL9    & 01 23 45.7 & -58 48 21.8 & 0.046  & 7.66 $\pm$ 0.09 & Quiet & FOS \\
    MRK 590     & 02 14 33.6 & -00 46 00.1 & 0.027  & 6.93 $\pm$ 0.06 & Quiet & IUE \\
    3C 120      & 04 33 11.1 & +05 21 15.6 & 0.033  & 7.00 $\pm$ 0.21\tablenotemark{c} & Loud & IUE \\
    ARK 120     & 05 16 11.4 & -00 08 59.4 & 0.033  & 7.43 $\pm$ 0.05 & Quiet & FOS \\ 
    MRK 79      & 07 42 32.8 & +49 48 34.7 & 0.0221 & 6.97 $\pm$ 0.11 & Quiet & IUE \\
    PG 0804+761 & 08 10 58.6 & +76 02 42.0 & 0.1000 & 8.10 $\pm$ 0.05 & Quiet & IUE \\
    MRK 110     & 09 25 12.9 & +52 17 10.5 & 0.0352 & 6.65 $\pm$ 0.10 & Quiet & IUE \\ 
    PG 0953+414 & 09 56 52.4 & +41 15 23.0 & 0.2341 & 7.69 $\pm$ 0.09 & Quiet & FOS \\ 
    NGC 3516    & 11 06 47.5 & +72 34 06.9 & 0.0088 & 6.88 $\pm$ 0.14 & Quiet & FOS \\
    NGC 3783    & 11 39 01.7 & -37 44 18.9 & 0.009  & 6.73 $\pm$ 0.07 & Quiet & FOS \\
    3C 273.0    & 12 29 06.7 & +02 03 09.0 & 0.1583 & 8.20 $\pm$ 0.09 & Loud  & FOS \\
    PG 1307+085 & 13 09 47.0 & +08 19 49.8 & 0.1550 & 7.90 $\pm$ 0.12 & Quiet & FOS \\
    MRK 279     & 13 53 03.4 & +69 18 29.6 & 0.0304 & 6.80 $\pm$ 0.26 & Quiet & STIS\\
    NGC 5548    & 14 17 59.5 & +25 08 12.4 & 0.0171 & 7.08 $\pm$ 0.01 & Quiet & FOS \\
    PG 1426+015 & 14 29 06.6 & +01 17 06.5 & 0.0864 & 8.37 $\pm$ 0.12 & Quiet & IUE \\
    MRK 817     & 14 36 22.1 & +58 47 39.4 & 0.0314 & 6.95 $\pm$ 0.06 & Quiet & IUE \\
    PG 1613+658 & 16 13 57.2 & +65 43 09.6 & 0.1290 & 7.70 $\pm$ 0.20\tablenotemark{c} & Quiet &  IUE \\
    PG 1617+175 & 16 20 11.3 & +17 24 27.7 & 0.1124 & 8.03 $\pm$ 0.10 & Quiet & IUE \\
    3C 390.3    & 18 42 08.8 & +79 46 17.0 & 0.0560 & 7.71 $\pm$ 0.09 & Loud  & FOS \\
    MRK 509     & 20 44 09.8 & -10 43 24.5 & 0.0350 & 7.41 $\pm$ 0.03 & Quiet & FOS \\
    PG 2130+099 & 21 32 27.8 & +10 08 19.5 & 0.0629 & 7.91 $\pm$ 0.05 & Quiet & GHRS\\
    NGC 7469    & 23 03 15.6 & +08 52 26.4 & 0.0163 & 6.34 $\pm$ 0.04 & Quiet & FOS \\
    \enddata
    
    \tablenotetext{a}{Black hole masses are from \citet{peter04},
    assuming $f = 1$.} \tablenotetext{b}{The radio type is either
    radio loud or radio quiet.}  \tablenotetext{c}{The standard errors
    in $M_{BH}$ for 3C 120 and PG 1613+658 were estimated by averaging
    the upper and lower uncertainties listed by \citet{peter04}.}
    
  \end{deluxetable}
  
  \clearpage
  
  \begin{deluxetable}{lcccccc}
    \tabletypesize{\scriptsize}
    \tablecaption{Log of New Observations\label{t-newobs}}
    \tablewidth{0pt}
    \tablehead{
      \colhead{Object} & \colhead{RA (J2000)} & \colhead{DEC (J2000)} & \colhead{Redshift} & 
      \colhead{Date} & \colhead{Instrument} & \colhead{Exp. Time (s)}
    }
    \startdata
    Q 0003+0146    & 00 05 47.6 & +02 03 02.2 & 0.234 & Oct 06, 2002 & Bok B\&C & 1800 \\
    Q 0017+0209    & 00 20 25.1 & +02 26 25.3 & 0.401 & Oct 06, 2002 & Bok B\&C & 2700 \\
    PKS 0044+030   & 00 47 05.9 & +03 19 54.9 & 0.624 & Oct 06, 2002 & Bok B\&C & 1800 \\
    Q 0100+0205    & 01 03 13.0 & +02 21 10.4 & 0.394 & Oct 06, 2002 & Bok B\&C & 2700 \\
    Q 0115+027     & 01 18 18.5 & +02 58 05.9 & 0.672 & Oct 06, 2002 & Bok B\&C & 3600 \\
    3C 57          & 02 01 57.2 & -11 32 33.7 & 0.669 & Oct 06, 2002 & Bok B\&C & 1800 \\
    PKS 0214+10    & 02 17 07.7 & +11 04 09.6 & 0.408 & Oct 06, 2002 & Bok B\&C & 2700 \\
    IR 0450-2958   & 04 52 30.0 & -29 53 35.0 & 0.286 & Oct 06, 2002 & Bok B\&C & 207  \\
    3C 232         & 09 58 21.0 & +32 24 02.2 & 0.533 & May 15, 2002 & Bok B\&C & 1800 \\
    4C 41.21       & 10 10 27.5 & +41 32 39.1 & 0.611 & May 15, 2002 & Bok B\&C & 1800 \\
    B2 1028+313    & 10 30 59.1 & +31 02 56.0 & 0.178 & May 15, 2002 & Bok B\&C & 1800 \\
    MC 1104+167    & 11 07 15.0 & +16 28 02.4 & 0.632 & May 15, 2002 & Bok B\&C & 1800 \\
    Q 1132-0302    & 11 35 04.9 & -03 18 52.5 & 0.237 & May 14, 2002 & Bok B\&C & 1800 \\
    Q 1150+497     & 11 53 24.5 & +49 31 08.6 & 0.334 & May 14, 2002 & Bok B\&C & 1200 \\
    TEX 1156+213   & 11 59 26.2 & +21 06 56.2 & 0.349 & May 15, 2002 & Bok B\&C & 2100 \\
    MRK 205        & 12 21 44.0 & +75 18 38.1 & 0.070 & May 15, 2002 & Bok B\&C & 1800 \\
    Q 1230+0947    & 12 33 25.8 & +09 31 23.0 & 0.420 & Mar 20, 2004 & Bok B\&C & 1800 \\
    Q 1317-0142    & 13 19 50.3 & -01 58 04.6 & 0.225 & May 14, 2002 & Bok B\&C & 900  \\
    PG 1333+176    & 13 36 02.0 & +17 25 13.0 & 0.554 & May 15, 2002 & Bok B\&C & 2100 \\
    Q 1340-0038    & 13 42 51.6 & -00 53 46.0 & 0.326 & May 14, 2002 & Bok B\&C & 1200 \\
    Q 1401+0951    & 14 04 10.6 & +09 37 45.5 & 0.441 & May 15, 2002 & Bok B\&C & 1800 \\
    Q 1425+2003    & 14 27 25.0 & +19 49 52.3 & 0.111 & May 14, 2002 & Bok B\&C & 900  \\
    PKS 1451-375   & 14 54 27.4 & -37 47 34.2 & 0.314 & Jul 17, 2004 & IMACS    & 2700 \\
    PG 1538+477    & 15 39 34.8 & +47 35 31.6 & 0.770 & May 14, 2002 & Bok B\&C & 1500 \\
    PKS 2128-12    & 21 31 35.4 & -12 07 05.5 & 0.501 & Oct. 06, 2002& Bok B\&C & 1800 \\
    PKS 2135-147   & 21 37 45.2 & -14 32 55.8 & 0.200 & May 15, 2002 & Bok B\&C & 900  \\
    Q 2141+175     & 21 43 35.6 & +17 43 49.1 & 0.213 & May 15, 2002 & Bok B\&C & 1200 \\
    PKS 2243-123   & 22 46 18.2 & -12 06 51.2 & 0.630 & Oct 06, 2002 & Bok B\&C & 1800 \\
    PKS 2352-342   & 23 55 25.6 & -33 57 55.8 & 0.706 & Jul 17, 2004 & IMACS    & 1800
    \enddata
    
  \end{deluxetable}
  
  \clearpage

  \begin{deluxetable}{lcccccc}
    \tabletypesize{\scriptsize}
    \tablecaption{Continuum and Fe Emission Fitting Windows\label{t-contwin}}
    \tablewidth{0pt}
    \tablehead{
      \colhead{} & \colhead{1} & \colhead{2} & \colhead{3} & \colhead{4} & \colhead{5} & \colhead{6}
    }
    \startdata
    UV      & $\lambda\lambda 1350$--$1365$ & $\lambda\lambda 1427$--$1500$ & $\lambda\lambda 1760$--$1860$ 
            & $\lambda\lambda 1950$--$2300$ & $\lambda\lambda 2470$--$2755$ & $\lambda\lambda 2855$--$3010$ \\
    Optical & $\lambda\lambda 3535$--$3700$ & $\lambda\lambda 4100$--$4200$ & $\lambda\lambda 4400$--$4700$ 
            & $\lambda\lambda 5100$--$6200$ & $\lambda\lambda 6800$--$7534$ & \nodata \\
    \enddata
  \end{deluxetable}

  \clearpage

  \begin{deluxetable}{lccccccc}
    \tabletypesize{\scriptsize}
    \tablecaption{Continuum and C IV Emission Line Parameters\label{t-params}}
    \tablewidth{0pt}
    \tablehead{
      \colhead{Object} & \colhead{$\log \lambda L_{\lambda} (1549\ \AA)$} &
      \colhead{$FWHM$} & \colhead{$\sigma_*$} & \colhead{$EW$} & \colhead{$\Delta v$} &
      \colhead{$\mu$} & \colhead{$\alpha$} \\
      \colhead{} & \colhead{ergs ${\rm cm}^{-2}$ ${\rm sec}^{-1}$} & 
      \colhead{$1000\ {\rm km\ s}^{-1}$} & \colhead{$1000\ {\rm km\ s}^{-1}$} & 
      \colhead{$\AA$} & \colhead{$1000\ {\rm km\ s}^{-1}$} & 
      \colhead{$1000\ {\rm km\ s}^{-1}$} & \colhead{}
    }
    \startdata
      MRK 335              & 44.38 & 2.094 & 1.856 & 71.82 & 0.368 & 0.141 & -0.45  \\
      PG0026+12            & 45.19 & 1.971 & 5.263 & 52.45 & 0.054 & -0.99 & -0.85  \\
      0052+2509            & 45.39 & 5.762 & 4.544 & 120.4 & -0.63 & -0.51 & -0.29  \\
      FAIRALL9             & 44.39 & 2.686 & 2.633 & 193.0 & -0.19 & -0.02 & -0.72  \\
      MRK 590              & 43.35 & 2.846 & 1.818 & 185.2 & -0.72 & -1.21 & -0.80  \\
      3C 120               & 44.56 & 5.040 & 2.813 & 148.3 & -1.34 & -0.87 & -0.77  \\
      ARK 120              & 43.91 & 4.127 & 3.124 & 70.0  & -0.00 & 0.647 & -0.95  \\
      MRK 79               & 43.68 & 4.615 & 2.891 & 112.4 & -0.41 & -0.68 & -0.73  \\
      PG 0804+761          & 45.26 & 3.608 & 3.307 & 62.63 & -0.41 & -1.20 & -0.79  \\
      MRK 110              & 43.77 & 3.468 & 2.117 & 185.0 & 0.819 & 0.437 & -0.79  \\
      PG0953+414           & 45.49 & 3.114 & 4.339 & 95.38 & -1.20 & 0.210 & -0.25  \\
      NGC 3516             & 42.31 & 3.512 & 2.480 & 327.9 & 0.054 & -0.47 & -2.05  \\
      NGC 3783             & 43.23 & 2.376 & 2.527 & 110.8 & 0.054 & -0.21 & -0.51  \\
      3C 273.0             & 46.28 & 3.659 & 3.285 & 31.24 & -1.70 & -0.64 & -0.57  \\
      1307+0835            & 45.08 & 4.054 & 3.395 & 96.97 & 0.431 & 0.339 & -0.35  \\
      MRK 279              & 44.32 & 3.057 & 3.201 & 110.2 & -0.85 & -1.00 & -0.29  \\
      NGC 5548             & 43.49 & 3.490 & 3.278 & 202.2 & 0.431 & 0.214 & -0.69  \\
      PG 1426+015          & 45.21 & 5.398 & 4.422 & 43.81 & -0.10 & -0.00 & -0.61  \\
      MRK 817              & 44.13 & 4.192 & 3.719 & 66.52 & -0.10 & 0.153 & -0.94  \\
      PG 1613+658          & 45.14 & 5.869 & 5.099 & 81.43 & -0.10 & 1.562 & -0.79  \\
      PG 1617+175          & 44.69 & 7.842 & 2.906 & 50.60 & 0.201 & 0.947 & -0.75  \\
      3C390.3              & 43.81 & 6.325 & 5.113 & 138.2 & -0.13 & -0.78 &  1.62  \\
      MRK509               & 44.68 & 4.313 & 3.722 & 128.3 & -0.95 & -0.13 & -0.77  \\
      PG 2130+099          & 44.60 & 2.200 & 1.785 & 58.36 & -0.78 & -0.94 & -0.78  \\
      NGC 7469             & 43.29 & 2.822 & 2.420 & 161.3 & -0.07 & 0.028 & -1.75  \\
      \enddata

  \end{deluxetable}

  \clearpage

  \begin{deluxetable}{cccccc}
    \tabletypesize{\scriptsize}
    \tablecaption{Black Hole Mass Estimates\label{t-mbhest}}
    \tablewidth{0pt}
    \tablehead{
      \colhead{RA (J2000)}
      & \colhead{DEC (J2000)}
      & \colhead{$z$}
      & \colhead{$\log M^{{\rm H}\beta}_{BH} / M_{\odot}$\tablenotemark{a}}
      & \colhead{$\log M^{CIV}_{BH} / M_{\odot}$}
      & \colhead{$\log M^{BL}_{BH} / M_{\odot}$\tablenotemark{b}}
    }
    \startdata
    00 05 47.6 & +02 03 02.2 & 0.234 & $ 7.32 \pm 0.43 $ & $ 7.61 \pm 0.40 $ & $ 7.47 \pm 0.29 $  \\
    00 05 59.2 & +16 09 49.1 & 0.450 & $ 8.38 \pm 0.43 $ & $ 8.18 \pm 0.40 $ & $ 8.27 \pm 0.29 $  \\
    00 06 19.5 & +20 12 10.3 & 0.025 & $ 6.72 \pm 0.43 $ & $ 6.90 \pm 0.40 $ & $ 6.82 \pm 0.29 $  \\
    00 20 25.1 & +02 26 25.3 & 0.401 & $ 7.23 \pm 0.43 $ & $ 8.33 \pm 0.40 $ & $ 7.82 \pm 0.29 $  \\
    00 29 13.7 & +13 16 03.8 & 0.142 & $ 7.30 \pm 0.43 $ & $ 7.17 \pm 0.40 $ & $ 7.23 \pm 0.29 $  \\
    00 47 05.9 & +03 19 54.9 & 0.624 & $ 9.07 \pm 0.43 $ & $ 8.67 \pm 0.40 $ & $ 8.86 \pm 0.29 $  \\
    00 52 02.4 & +01 01 29.3 & 2.270 & $ 9.47 \pm 0.44 $ & $ 9.41 \pm 0.40 $ & $ 9.44 \pm 0.29 $  \\
    00 52 33.7 & +01 40 40.6 & 2.307 & $ 9.25 \pm 0.46 $ & $ 9.55 \pm 0.40 $ & $ 9.42 \pm 0.30 $  \\
    00 54 52.1 & +25 25 39.3 & 0.155 & $ 7.89 \pm 0.43 $ & $ 8.34 \pm 0.40 $ & $ 8.13 \pm 0.29 $  \\
    01 03 13.0 & +02 21 10.4 & 0.394 & $ 8.47 \pm 0.43 $ & $ 8.16 \pm 0.40 $ & $ 8.30 \pm 0.29 $  \\
    01 18 18.5 & +02 58 05.9 & 0.672 & $ 8.38 \pm 0.44 $ & $ 7.61 \pm 0.40 $ & $ 7.96 \pm 0.29 $  \\
    01 23 45.7 & -58 48 21.8 & 0.047 & $ 7.40 \pm 0.43 $ & $ 7.16 \pm 0.40 $ & $ 7.27 \pm 0.29 $  \\
    01 26 42.8 & +25 59 01.3 & 2.370 & $ 9.24 \pm 0.44 $ & $ 9.25 \pm 0.40 $ & $ 9.25 \pm 0.30 $  \\
    01 57 35.0 & +74 42 43.2 & 2.338 & $ 9.10 \pm 0.44 $ & $ 9.20 \pm 0.41 $ & $ 9.15 \pm 0.30 $  \\
    02 01 57.2 & -11 32 33.7 & 0.669 & $ 8.01 \pm 0.47 $ & $ 8.85 \pm 0.40 $ & $ 8.49 \pm 0.30 $  \\
    02 17 07.7 & +11 04 09.6 & 0.408 & $ 8.69 \pm 0.43 $ & $ 8.20 \pm 0.40 $ & $ 8.42 \pm 0.29 $  \\
    02 59 05.6 & +00 11 21.9 & 3.366 & $ 8.95 \pm 0.43 $ & $ 9.03 \pm 0.40 $ & $ 8.99 \pm 0.29 $  \\
    03 04 49.9 & -00 08 13.4 & 3.294 & $ 8.90 \pm 0.43 $ & $ 8.69 \pm 0.40 $ & $ 8.79 \pm 0.29 $  \\
    03 51 28.6 & -14 29 09.1 & 0.616 & $ 8.92 \pm 0.43 $ & $ 9.02 \pm 0.40 $ & $ 8.98 \pm 0.29 $  \\
    04 05 34.0 & -13 08 14.1 & 0.571 & $ 8.56 \pm 0.43 $ & $ 8.05 \pm 0.40 $ & $ 8.29 \pm 0.29 $  \\
    04 07 48.4 & -12 11 36.0 & 0.574 & $ 8.25 \pm 0.43 $ & $ 8.41 \pm 0.40 $ & $ 8.34 \pm 0.29 $  \\
    04 17 16.8 & -05 53 45.9 & 0.781 & $ 9.04 \pm 0.45 $ & $ 8.96 \pm 0.40 $ & $ 9.00 \pm 0.30 $  \\
    04 41 17.3 & -43 13 43.7 & 0.593 & $ 8.08 \pm 0.43 $ & $ 7.84 \pm 0.40 $ & $ 7.95 \pm 0.29 $  \\
    04 52 30.0 & -29 53 35.0 & 0.286 & $ 7.37 \pm 0.43 $ & $ 7.86 \pm 0.40 $ & $ 7.64 \pm 0.29 $  \\
    04 56 08.9 & -21 59 09.4 & 0.534 & $ 8.73 \pm 0.43 $ & $ 8.49 \pm 0.40 $ & $ 8.60 \pm 0.29 $  \\
    07 45 41.7 & +31 42 55.7 & 0.462 & $ 8.27 \pm 0.43 $ & $ 8.54 \pm 0.40 $ & $ 8.41 \pm 0.29 $  \\
    08 40 47.6 & +13 12 23.7 & 0.684 & $ 8.06 \pm 0.43 $ & $ 8.09 \pm 0.40 $ & $ 8.07 \pm 0.29 $  \\
    08 53 34.2 & +43 49 01.0 & 0.513 & $ 7.84 \pm 0.43 $ & $ 8.41 \pm 0.40 $ & $ 8.15 \pm 0.29 $  \\
    09 19 57.7 & +51 06 10.0 & 0.553 & $ 8.26 \pm 0.43 $ & $ 8.58 \pm 0.40 $ & $ 8.43 \pm 0.29 $  \\
    09 50 48.4 & +39 26 51.0 & 0.206 & $ 8.53 \pm 0.43 $ & $ 7.88 \pm 0.40 $ & $ 8.18 \pm 0.29 $  \\
    09 56 52.4 & +41 15 23.0 & 0.239 & $ 7.70 \pm 0.43 $ & $ 7.85 \pm 0.40 $ & $ 7.78 \pm 0.29 $  \\
    09 58 21.0 & +32 24 02.2 & 0.533 & $ 7.48 \pm 0.43 $ & $ 8.87 \pm 0.40 $ & $ 8.22 \pm 0.30 $  \\
    10 04 02.6 & +28 55 35.0 & 0.329 & $ 6.86 \pm 0.43 $ & $ 8.25 \pm 0.40 $ & $ 7.61 \pm 0.29 $  \\
    10 04 20.1 & +05 13 00.0 & 0.161 & $ 7.60 \pm 0.43 $ & $ 7.40 \pm 0.40 $ & $ 7.49 \pm 0.29 $  \\
    10 10 27.5 & +41 32 39.1 & 0.611 & $ 7.99 \pm 0.43 $ & $ 8.38 \pm 0.40 $ & $ 8.20 \pm 0.29 $  \\
    10 30 59.1 & +31 02 56.0 & 0.178 & $ 7.97 \pm 0.44 $ & $ 7.59 \pm 0.40 $ & $ 7.76 \pm 0.30 $  \\
    10 51 51.5 & +00 51 18.1 & 0.357 & $ 7.88 \pm 0.43 $ & $ 8.18 \pm 0.40 $ & $ 8.04 \pm 0.29 $  \\
    11 04 13.9 & +76 58 58.2 & 0.311 & $ 8.44 \pm 0.43 $ & $ 7.92 \pm 0.40 $ & $ 8.16 \pm 0.29 $  \\
    11 06 31.8 & +00 52 53.4 & 0.425 & $ 8.67 \pm 0.43 $ & $ 8.43 \pm 0.40 $ & $ 8.54 \pm 0.29 $  \\
    11 06 33.5 & -18 21 24.0 & 2.319 & $ 8.89 \pm 0.45 $ & $ 9.65 \pm 0.40 $ & $ 9.31 \pm 0.30 $  \\
    11 07 15.0 & +16 28 02.4 & 0.632 & $ 8.04 \pm 0.45 $ & $ 8.65 \pm 0.40 $ & $ 8.38 \pm 0.30 $  \\
    11 18 30.3 & +40 25 55.0 & 0.154 & $ 7.82 \pm 0.43 $ & $ 7.93 \pm 0.40 $ & $ 7.88 \pm 0.29 $  \\
    11 19 08.7 & +21 19 18.0 & 0.176 & $ 7.55 \pm 0.43 $ & $ 8.13 \pm 0.40 $ & $ 7.87 \pm 0.29 $  \\
    11 24 39.2 & +42 01 45.2 & 0.234 & $ 7.61 \pm 0.43 $ & $ 7.47 \pm 0.40 $ & $ 7.53 \pm 0.29 $  \\
    11 35 04.9 & -03 18 52.5 & 0.237 & $ 7.07 \pm 0.43 $ & $ 7.87 \pm 0.40 $ & $ 7.50 \pm 0.29 $  \\
    11 39 57.1 & +65 47 49.4 & 0.652 & $ 7.60 \pm 0.43 $ & $ 8.22 \pm 0.40 $ & $ 7.93 \pm 0.29 $  \\
    11 41 21.7 & +01 48 03.3 & 0.383 & $ 7.20 \pm 0.45 $ & $ 8.23 \pm 0.40 $ & $ 7.77 \pm 0.30 $  \\
    11 47 18.0 & -01 32 07.7 & 0.382 & $ 8.08 \pm 0.43 $ & $ 6.68 \pm 0.43 $ & $ 7.37 \pm 0.30 $  \\
    11 53 24.5 & +49 31 08.6 & 0.334 & $ 7.68 \pm 0.43 $ & $ 8.02 \pm 0.40 $ & $ 7.86 \pm 0.29 $  \\
    11 58 39.9 & +62 54 28.1 & 0.594 & $ 7.45 \pm 0.51 $ & $ 8.32 \pm 0.40 $ & $ 7.99 \pm 0.32 $  \\
    11 59 26.2 & +21 06 56.2 & 0.349 & $ 8.69 \pm 0.43 $ & $ 8.03 \pm 0.40 $ & $ 8.33 \pm 0.29 $  \\
    12 04 42.2 & +27 54 12.0 & 0.165 & $ 8.24 \pm 0.43 $ & $ 7.26 \pm 0.40 $ & $ 7.71 \pm 0.29 $  \\
    12 14 17.7 & +14 03 12.3 & 0.080 & $ 6.72 \pm 0.43 $ & $ 6.95 \pm 0.40 $ & $ 6.84 \pm 0.29 $  \\
    12 19 20.9 & +06 38 38.4 & 0.334 & $ 8.00 \pm 0.43 $ & $ 7.95 \pm 0.40 $ & $ 7.97 \pm 0.29 $  \\
    12 21 44.0 & +75 18 38.1 & 0.070 & $ 7.89 \pm 0.43 $ & $ 7.59 \pm 0.40 $ & $ 7.73 \pm 0.29 $  \\
    12 31 20.6 & +07 25 52.8 & 2.391 & $ 9.14 \pm 0.49 $ & $ 9.19 \pm 0.40 $ & $ 9.17 \pm 0.31 $  \\
    12 33 25.8 & +09 31 23.0 & 0.420 & $ 7.98 \pm 0.43 $ & $ 8.18 \pm 0.40 $ & $ 8.09 \pm 0.29 $  \\
    13 01 12.9 & +59 02 06.9 & 0.472 & $ 7.91 \pm 0.43 $ & $ 9.01 \pm 0.40 $ & $ 8.50 \pm 0.29 $  \\
    13 05 33.0 & -10 33 20.4 & 0.286 & $ 8.19 \pm 0.43 $ & $ 8.38 \pm 0.40 $ & $ 8.29 \pm 0.29 $  \\
    13 09 47.0 & +08 19 49.8 & 0.155 & $ 8.54 \pm 0.43 $ & $ 7.87 \pm 0.40 $ & $ 8.18 \pm 0.29 $  \\
    13 12 17.7 & +35 15 21.0 & 0.184 & $ 7.62 \pm 0.43 $ & $ 7.43 \pm 0.40 $ & $ 7.52 \pm 0.29 $  \\
    13 19 50.3 & -01 58 04.6 & 0.225 & $ 7.05 \pm 0.43 $ & $ 7.23 \pm 0.40 $ & $ 7.15 \pm 0.29 $  \\
    13 23 49.5 & +65 41 48.0 & 0.168 & $ 7.11 \pm 0.43 $ & $ 7.60 \pm 0.40 $ & $ 7.37 \pm 0.29 $  \\
    13 36 02.0 & +17 25 13.0 & 0.554 & $ 7.72 \pm 0.43 $ & $ 8.79 \pm 0.40 $ & $ 8.29 \pm 0.29 $  \\
    13 42 51.6 & -00 53 46.0 & 0.326 & $ 8.27 \pm 0.43 $ & $ 8.20 \pm 0.40 $ & $ 8.23 \pm 0.29 $  \\
    13 57 04.5 & +19 19 06.6 & 0.719 & $ 7.12 \pm 0.44 $ & $ 7.99 \pm 0.40 $ & $ 7.60 \pm 0.30 $  \\
    14 04 10.6 & +09 37 45.5 & 0.441 & $ 9.06 \pm 0.45 $ & $ 8.49 \pm 0.41 $ & $ 8.74 \pm 0.30 $  \\
    14 05 16.2 & +25 55 33.6 & 0.164 & $ 7.35 \pm 0.43 $ & $ 8.19 \pm 0.40 $ & $ 7.80 \pm 0.29 $  \\
    14 17 00.9 & +44 56 06.0 & 0.114 & $ 7.61 \pm 0.43 $ & $ 7.68 \pm 0.40 $ & $ 7.65 \pm 0.29 $  \\
    14 19 03.9 & -13 10 45.0 & 0.129 & $ 7.56 \pm 0.43 $ & $ 7.76 \pm 0.40 $ & $ 7.67 \pm 0.29 $  \\
    14 27 25.0 & +19 49 52.3 & 0.111 & $ 8.03 \pm 0.43 $ & $ 8.30 \pm 0.40 $ & $ 8.17 \pm 0.29 $  \\
    14 27 35.7 & +26 32 15.0 & 0.366 & $ 7.85 \pm 0.43 $ & $ 8.80 \pm 0.40 $ & $ 8.36 \pm 0.29 $  \\
    14 29 43.1 & +47 47 26.0 & 0.221 & $ 7.65 \pm 0.43 $ & $ 7.64 \pm 0.40 $ & $ 7.64 \pm 0.29 $  \\
    14 42 07.5 & +35 26 22.9 & 0.077 & $ 7.01 \pm 0.43 $ & $ 7.26 \pm 0.40 $ & $ 7.15 \pm 0.29 $  \\
    14 46 45.9 & +40 35 07.1 & 0.267 & $ 7.38 \pm 0.43 $ & $ 8.42 \pm 0.40 $ & $ 7.94 \pm 0.29 $  \\
    14 54 27.4 & -37 47 34.2 & 0.314 & $ 7.62 \pm 0.43 $ & $ 7.93 \pm 0.40 $ & $ 7.79 \pm 0.29 $  \\
    15 14 43.5 & +36 50 51.0 & 0.371 & $ 7.84 \pm 0.44 $ & $ 8.28 \pm 0.40 $ & $ 8.08 \pm 0.30 $  \\
    15 39 34.8 & +47 35 31.6 & 0.770 & $ 8.58 \pm 0.46 $ & $ 9.01 \pm 0.40 $ & $ 8.82 \pm 0.30 $  \\
    15 45 30.3 & +48 46 07.9 & 0.400 & $ 7.73 \pm 0.43 $ & $ 8.55 \pm 0.40 $ & $ 8.17 \pm 0.29 $  \\
    15 47 43.5 & +20 52 16.4 & 0.264 & $ 8.39 \pm 0.43 $ & $ 8.21 \pm 0.40 $ & $ 8.29 \pm 0.29 $  \\
    16 14 13.2 & +26 04 16.2 & 0.131 & $ 7.05 \pm 0.43 $ & $ 7.78 \pm 0.40 $ & $ 7.44 \pm 0.29 $  \\
    16 20 21.8 & +17 36 24.0 & 0.555 & $ 8.22 \pm 0.43 $ & $ 8.69 \pm 0.40 $ & $ 8.47 \pm 0.29 $  \\
    16 27 56.1 & +55 22 31.0 & 0.133 & $ 8.49 \pm 0.43 $ & $ 7.89 \pm 0.40 $ & $ 8.17 \pm 0.29 $  \\
    16 42 58.8 & +39 48 36.9 & 0.595 & $ 8.01 \pm 0.43 $ & $ 8.45 \pm 0.40 $ & $ 8.24 \pm 0.29 $  \\
    17 04 41.3 & +60 44 30.0 & 0.371 & $ 7.37 \pm 0.44 $ & $ 8.15 \pm 0.40 $ & $ 7.80 \pm 0.30 $  \\
    18 21 59.4 & +64 21 07.5 & 0.297 & $ 8.85 \pm 0.43 $ & $ 8.53 \pm 0.40 $ & $ 8.68 \pm 0.29 $  \\
    19 27 48.5 & +73 58 02.0 & 0.302 & $ 8.13 \pm 0.43 $ & $ 8.19 \pm 0.40 $ & $ 8.16 \pm 0.29 $  \\
    20 44 09.8 & -10 43 24.5 & 0.035 & $ 7.64 \pm 0.43 $ & $ 7.69 \pm 0.40 $ & $ 7.66 \pm 0.29 $  \\
    21 31 35.4 & -12 07 05.5 & 0.501 & $ 8.13 \pm 0.43 $ & $ 8.64 \pm 0.40 $ & $ 8.40 \pm 0.29 $  \\
    21 37 45.2 & -14 32 55.8 & 0.200 & $ 8.86 \pm 0.43 $ & $ 7.99 \pm 0.43 $ & $ 8.42 \pm 0.31 $  \\
    21 43 35.6 & +17 43 49.1 & 0.213 & $ 8.24 \pm 0.43 $ & $ 8.20 \pm 0.41 $ & $ 8.22 \pm 0.30 $  \\
    22 03 15.0 & +31 45 37.7 & 0.297 & $ 8.08 \pm 0.43 $ & $ 8.76 \pm 0.40 $ & $ 8.45 \pm 0.29 $  \\
    22 46 18.2 & -12 06 51.2 & 0.630 & $ 8.41 \pm 0.43 $ & $ 8.13 \pm 0.40 $ & $ 8.26 \pm 0.29 $  \\
    22 54 05.8 & -17 34 55.0 & 0.068 & $ 8.15 \pm 0.43 $ & $ 7.56 \pm 0.40 $ & $ 7.83 \pm 0.29 $  \\
    22 54 10.4 & +11 36 38.9 & 0.323 & $ 7.59 \pm 0.43 $ & $ 8.13 \pm 0.40 $ & $ 7.88 \pm 0.29 $  \\
    23 03 43.5 & -68 07 37.1 & 0.512 & $ 7.62 \pm 0.43 $ & $ 7.77 \pm 0.40 $ & $ 7.70 \pm 0.29 $  \\
    23 11 17.8 & +10 08 16.2 & 0.432 & $ 8.48 \pm 0.43 $ & $ 8.36 \pm 0.40 $ & $ 8.42 \pm 0.29 $  \\
    23 46 36.9 & +09 30 46.0 & 0.672 & $ 7.94 \pm 0.43 $ & $ 7.99 \pm 0.40 $ & $ 7.97 \pm 0.29 $  \\
    23 51 56.0 & -01 09 13.7 & 0.174 & $ 8.52 \pm 0.43 $ & $ 8.09 \pm 0.40 $ & $ 8.29 \pm 0.29 $  \\
    23 55 25.6 & -33 57 55.8 & 0.706 & $ 6.51 \pm 0.44 $ & $ 8.39 \pm 0.40 $ & $ 7.53 \pm 0.30 $  \\
    \enddata
    \tablecomments{The $1\sigma$ uncertainties include the
      contributions from measurement error in the line widths and
      continuum luminosities, and from the intrinsic uncertainty in
      $\hat{m}$.}
    \tablenotetext{a}{The H$\beta$-based mass estimates are for the UV
      $R_{{\rm H}\beta}$--$L$ relationship. The $1\sigma$ errors were
      calculated assuming $\sigma_{{\rm H}\beta} = 0.43$ dex
      \citep{vest06}.}
    \tablenotetext{b}{$M^{BL}_{BH}$ is a weighted average of the
      H$\beta$- and C IV-based mass estimates.}

  \end{deluxetable}

\end{document}